%% file: paper.tex
\PassOptionsToPackage{hyphens}{url}
\documentclass[sigconf]{acmart}

\input{packages}
\input{listing-setup}

\copyrightyear{2025}
\acmYear{2025}
\setcopyright{cc}
\setcctype{by}
\acmConference[CCS '25]{Proceedings of the 2025 ACM SIGSAC Conference on Computer and Communications Security}{October 13--17, 2025}{Taipei, Taiwan}
\acmBooktitle{Proceedings of the 2025 ACM SIGSAC Conference on Computer and Communications Security (CCS '25), October 13--17, 2025, Taipei, Taiwan}
\acmDOI{10.1145/3719027.3765195}
\acmISBN{979-8-4007-1525-9/2025/10}

\begin{document}
	\title[Passwords and FIDO2 Are Meant To Be Secret: A Practical Secure Authentication Channel for Web Browsers]{Passwords and FIDO2 Are Meant To Be Secret:\\A Practical Secure Authentication Channel for Web Browsers}
	
	\author{Anuj Gautam}
	\affiliation{
		\institution{University of Tennessee}
		\city{Knoxville}
		\state{TN}
		\country{USA}
	}
	\email{agautam1@alum.utk.edu}
	
	\author{Tarun Yadav}
	\affiliation{
		\institution{Brigham Young University}
		\city{Provo}
		\state{UT}
		\country{USA}
	}
	\email{tarun14110@iiitd.ac.in}
	
	\author{Garrett Smith}
	\affiliation{
		\institution{Brigham Young University}
		\city{Provo}
		\state{UT}
		\country{USA}
	}
	\email{gds33@byu.edu}
	
	\author{Kent Seamons}
	\affiliation{
		\institution{Brigham Young University}
		\city{Provo}
		\state{UT}
		\country{USA}
	}
	\email{seamons@cs.byu.edu}
	
	\author{Scott Ruoti}
	\affiliation{
		\institution{University of Tennessee}
		\city{Knoxville}
		\state{TN}
		\country{USA}
	}
	\email{ruoti@utk.edu}
	
	\renewcommand{\shortauthors}{Anuj Gautam, Tarun Yadav, Garrett Smith, Kent Seamons, \& Scott Ruoti}
	
	\input{abstract}



	\maketitle
	
	%-------------------------------------------------------------------------------
	\input{introduction}
	\input{background}
	\input{threats}
	\input{sbc-pwm}
	\input{threats-full}
	\input{sbc-pwm-evaluation}
	\input{porting}
	\input{sbc-2fa}
	\input{discussion}
	\input{rw}
	\input{conclusion}

	\section*{Acknowledgments}
	This material is based upon work supported by the National Science Foundation under Grants No. CNS-2226404 and CNS-1816929. Any opinions, findings, and conclusions or recommendations expressed in this material are those of the authors and do not necessarily reflect the views of the National Science Foundation.
	
	%-------------------------------------------------------------------------------
	
	\bibliographystyle{ACM-Reference-Format}
	\balance
	\bibliography{authentication,publications,paper,2fa}

 	arXiv only
	\clearpage
	\appendix
	\input{appx-threats}
	\input{appx-design}
\end{document}

%% file: packages.tex
\usepackage{formatting}
\usepackage{tables}

\usepackage{tikz}
\usepackage{pgf-umlsd} 
\usepackage{hyperref}

%% file: listing-setup.tex
\usepackage{listings}
\usepackage{color}
\usepackage{textcomp}
\usepackage{lstautogobble} 

\usepackage[T1]{fontenc}

\lstdefinelanguage{json}
{
	morestring=[b]",
	morestring=[d]',
	morecomment=[s]{/*}{*/},
	xleftmargin={0pt},
}

\lstdefinelanguage{JavaScript}{
	morekeywords=[1]{break, continue, delete, else, for, function, if, in,
		new, return, this, typeof, var, void, while, with},
	morekeywords=[2]{false, null, true, boolean, number, undefined,
		Array, Boolean, Date, Math, Number, String, Object},
	morekeywords=[3]{eval, parseInt, parseFloat, escape, unescape},
	sensitive,
	morecomment=[s]{/*}{*/},
	morecomment=[l]//,
	morecomment=[s]{/**}{*/}, % JavaDoc style comments
	morestring=[b]',
	morestring=[b]"
}[keywords, comments, strings]

\lstdefinelanguage[ECMAScript2015]{JavaScript}[]{JavaScript}{
	morekeywords=[1]{await, async, case, catch, class, const, default, do,
		enum, export, extends, finally, from, implements, import, instanceof,
		let, static, super, switch, throw, try},
	morestring=[b]` % Interpolation strings.
}

\lstalias[]{ES6}[ECMAScript2015]{JavaScript}

\definecolor{mediumgray}{rgb}{0.3, 0.4, 0.4}
\definecolor{mediumblue}{rgb}{0.0, 0.0, 0.8}
\definecolor{forestgreen}{rgb}{0.13, 0.55, 0.13}
\definecolor{darkviolet}{rgb}{0.58, 0.0, 0.83}
\definecolor{royalblue}{rgb}{0.25, 0.41, 0.88}
\definecolor{crimson}{rgb}{0.86, 0.8, 0.24}

\definecolor{backgroundColor}{HTML}{ffffff} 		% fg.default
\definecolor{identifierColor}{HTML}{1F2328} 		% canvas.default
\definecolor{commentColor}{HTML}{6e7781}			% prettylights.comment
\definecolor{stringColor}{HTML}{0a3069}				% prettylights.string
\definecolor{keywordColor}{HTML}{cf222e}			% prettylights.keyword
\definecolor{valueKeywordColor}{HTML}{0550ae}		% prettylights.constant
\definecolor{functionKeywordColor}{HTML}{8250df}	% prettylights.entity
\definecolor{emphasisColor}{HTML}{82071e}			% prettylights.brackethighlighterUnmatched

\lstdefinestyle{base}{
	autogobble,
	showlines=true,
	emptylines=1,
	basicstyle=\ttfamily\scriptsize,
	identifierstyle=\color{identifierColor},
	commentstyle=\color{commentColor}\upshape,
	stringstyle=\color{stringColor},
	keywordstyle=\color{keywordColor},
	keywordstyle=[2]\color{valueKeywordColor},
	keywordstyle=[3]\color{functionKeywordColor},
	emphstyle=\color{emphasisColor},
	extendedchars=true,
	upquote=true,
	tabsize=2,
	showtabs=false,
	showspaces=false,
	showstringspaces=false,
	numbers=left,
	numberstyle=\scriptsize\color{black},
	numbersep=5pt,
	numberblanklines=true,
	captionpos=b,
	xleftmargin=2em,
	xrightmargin=1em,
	frame=single,
	backgroundcolor=\color{backgroundColor},
}

%% file: abstract.tex
%!TEX root=paper.tex

\begin{abstract}
%Password-based authentication faces various security and usability issues.
Password managers provide significant security benefits to users.
However, malicious client-side scripts and browser extensions can steal passwords after the manager has autofilled them into the web page.
In this paper, we extend prior work by Stock and Johns, showing how password autofill can be hardened to prevent these local attacks.
We implement our design in the Firefox browser and conduct experiments demonstrating that our defense successfully protects passwords from XSS attacks and malicious extensions.
We also show that our implementation is compatible with 97\% of the Alexa top 1000 websites.
% while also maintaining the capability to revert to default behavior on the remaining websites, avoiding functionality regressions.
%As such, our work represents a significant step forward in the security that password managers can provide for password-based authentication in the browser.
Next, we generalize our design, creating a second defense that prevents recently discovered local attacks against the FIDO2 protocols.
%Most importantly, both designs is transparent to users, requiring no change to user behavior.
We implement this second defense into Firefox, demonstrating that it protects the FIDO2 protocol against XSS attacks and malicious extensions.
This defense is compatible with all websites, though it does require a small change (2--3 lines) to web servers implementing FIDO2.
%Ultimately, this work represents both a scientific and practical contribution to password security.
%Finally, we conclude with a discussion of future research directions that could generalize and further secure the proposed design.
\end{abstract}

%% file: introduction.tex
%!TEX root=paper.tex

\section{Introduction}

%Despite their many problems~\cite{florencio2007large,dellamico2010password,ur2015added,pearman2017lets}, passwords remain the dominant form of authentication on the Web.
%Even with the rise of two-factor authentication (2FA), passwords are usually included as one of the two factors.
%While passwords may someday be fully replaced by more secure alternatives~\cite{bonneau2012quest}, for now, there is an urgent need to secure the use of passwords.

Password managers seek to improve the security of passwords by helping users generate, store, and enter strong passwords~\cite{oesch2020that,simmons2021systematization}.
However, password manager security has a conspicuous gap---passwords are vulnerable to theft after they are autofilled into the browser and before they are transmitted to the website.
This includes theft from web trackers~\cite{senol2022leaky}, injection attacks (e.g., XSS)~\cite{help_net_security_2021}, malicious browser extensions~\cite{kapravelos2014hulk,pantelaios2020you,kaspersky2023dangerous}, or compromised JavaScript libraries~\cite{ohm2020backstabber,duan2020towards} (i.e., a supply chain attack).
With the relative widespread nature of these issues~\cite{xssowasp,CVEdetails2024,sansec2024polyfill}, this security gap urgently needs to be closed.
%As such, this paper explores what can be done to address this gap with minimal impact on users and websites.

A decade ago, Stock and Johns~\cite{stock2014protecting} proposed modifying password managers so that instead of autofilling the user's password, the extension would fill a random value (i.e., a nonce).
Then, after the page is submitted but before the related HTTP request is transmitted over the network, the nonce is replaced with the actual password.
Critically, the actual password is never visible to the webpage, preventing DOM-based attacks.
%There are two significant limitations to this approach.
%First, it does not work in modern browsers, which prevents extensions from modifying the body of HTTP requests.
%Second, it does not protect against malicious or compromised browser extensions.
Unfortunately, even at the time this protocol was proposed, it was incompatible with modern browser design~\cite{chromium91191webrequest}, and the one browser where it did work, Firefox, removed this functionality shortly after~\cite{mozilla1376155webrequest}.
%Moreover, there was no obvious path for getting this protocol to work in browsers, short of the daunting task of figuring out how to modify the browsers themselves.

%In this paper, we create a new browser-based system design that addresses both of these limitations.
%To demonstrate the feasibility of this design, we forked and modified the Firefox browser and Bitwarden, an open-source password manager, to implement our design.
%%We also modified BitWarden, an open-source password manager, to work with our modified Firefox implementation.
%%Importantly, our proof-of-concept implementation does not require modifying user behavior or websites.
%We empirically evaluated the security of our design, demonstrating that it stops password exfiltration by DOM- and extension-based adversaries.
%Moreover, we describe how to prevent attacks by adversaries who are aware of this defense and may try to subvert it.
%We also evaluated our tool on 573 sites with login forms from the Alexa Top 1000, showing that it is compatible with 97\% of those sites.
%For the remaining 3\% of websites, the password manager can easily revert to its existing behavior, preventing any functionality regression.
%\emph{This represents an immediate and significant step forward for the security of password-based authentication when using a password manager.}

Addressing this limitation, this paper's first contribution is practical---we describe how modern browsers can be modified to add a nonce-based password replacement API.
Our API limits the extent to which browser requests can be modified and provides safeguards that allow password managers to ensure passwords are sent only to the appropriate destination.
To demonstrate the feasibility of this design, we forked and modified the Firefox browser to implement our API and Bitwarden to use it.
We empirically evaluated the security of our implementations, demonstrating that they stop password exfiltration by DOM-based adversaries (e.g., web trackers, XSS, compromised JavaScript libraries).
We also evaluated our API on 573 sites with login forms from the Alexa Top 1000, showing that it is compatible with 97\% of those sites.
%For the remaining 3\% of websites, the password manager can easily revert to its existing behavior, preventing any functionality regression.

%Implementing this API was a significant engineering effort, requiring hundreds of hours of research and development.
Our work has substantial practical and research implications.
First, it represents an immediate and significant step forward for the security of password-based authentication for tens of millions of password manager users.
Second, it allows the research community to build on and expand both our API and the protocol proposed by Stock and Johns.
In the decade since Stock and Johns' original paper, there has been no research expanding on this promising work~\cite{oesch2020that}, likely because it was non-functional.
By providing a functional implementation of Stock and Johns' protocol on modern browsers, researchers can explore how to extend or use our API to further secure passwords (see \S\ref{sec:fw}) or other browser-server operations (see \S\ref{sec:additional-applications}).

Our second contribution is that we identify a gap in Stock and Johns's original threat model: malicious browser extensions that can exfiltrate the password by examining outgoing HTTP requests.
In an analysis of extensions on the Chrome web store, we identified thousands of extensions with sufficient permissions to conduct such an attack.
We then demonstrate how our defense can be modified to prevent this new class of attacks.
%We designed our API to be resistant to this attack, providing a tangible improvement to Stock and Johns' original protocol.

Inspired by our API, we hypothesized that similar APIs could be used to harden other browser-server services.
Our paper's third contribution is that we prove this hypothesis correct by developing a second defense that secures FIDO2 against recently discovered local attacks---i.e., DOM-based attacks (e.g., XSS, compromised JavaScript libraries) and malicious extensions~\cite{guan2022formal,hu2016security,yadav2024security,mahdad2024breaching}.
This second API provides a confidential channel for the web server and the client's FIDO2 device to communicate without allowing the web page's DOM or browser extensions to access sensitive values.
This design requires no changes for FIDO2 devices and only a trivial change (2--3 lines) to the web servers implementing the FIDO2 protocol.
We implement this API into Firefox and evaluate it empirically.
%As with our first design, we demonstrate its feasibility by implementing it into Firefox.

%TODO: Is this enough? Does it ever go to far?
%The contributions of this paper are both theoretical and practical.
%Theoretically, we provide a rigorous threat model for password exfiltration, including measurements showing the large number of extensions that could potentially steal credentials if compromised.
%We then demonstrate how Stock and Johns' work~\cite{stock2014protecting} can be extended to defend against this more rigorous threat model.
%We also generalize our defense, demonstrating how it can be used to also protect from recently discovered local attacks against the FIDO2 protocol.
%Practically, we demonstrate how these defenses can be implemented in modern browsers, making this code available to the community through the Firefox repository (branch names redacted for anonymity).
%Due to the complexity of modern browsers, this contribution required several hundred hours of work and consultation with the Firefox team to ensure that our defense did not compromise functionality or security within the browser.

%% file: background.tex
%!TEX root=paper.tex

\section{Background}

In this section, we provide background material on the password autofill workflow, how browsers process forms and web requests, a DOM-centric threat model for password exfiltration, and Stock and Johns' proposal for addressing DOM-based password exfiltration~\cite{stock2014protecting}.

%Below, we describe the technical details of entering a password in the browser, how form submission is handled, and how browser extensions can interact with web requests.

\subsection{Password Autofill Workflow} \label{sec:workflow}
The password autofill workflow includes the following steps:

\begin{enumerate}
	\item The user visits a webpage with a \texttt{form} containing an \texttt{input} element to enter their password.
	\item The password manager detects this form and offers to autofill the password. It does so by providing the user with a list of passwords associated with the current domain that they can autofill.
	\item The user selects which password to autofill.\footnote{This process is automated in some password managers, though it is unsafe to do so~\cite{oesch2020that}.}
	\item The password manager sets the \texttt{value} field of the password \texttt{input} element. At this point, the password is stored within the web page's DOM.
	\item The user submits the form.
	\item The browser creates a \texttt{webRequest} object responsible for sending the password to the server.
	\item The password is transferred over the network by the browser (preferably using a TLS connection).
\end{enumerate}

This flow also applies to account creation, with the same general process for entering and transmitting passwords.

\subsection{Browser Functionality}

\subsubsection{Browser Form Submission}
\label{sec:formsubmission}

The form submission process is defined in the HTML specifications~\cite{formsubmission} and starts when the \texttt{submit()} method is called on a \texttt{FormElement} object.
This can happen when the user clicks a submit button associated with that form or when JavaScript directly calls this method.

First, the browser will construct an entry list consisting of all the input fields in the form~\cite{html5constructentrylist}.
This process involves sanitizing textual inputs and field names and converting checkboxes and radio selections into strings representing the selected items.
Next, the form data is encoded for network transmission based on the form's \texttt{accept-charset} attribute.
The browser will then convert all values into name-value pairs, applying URL encoding if required by the \texttt{accept-charset} attribute.
These items are all put into \texttt{FormSubmission} object.
Finally, the browser will create a \texttt{webRequest} object that sends a web request to the server.
The web request's destination is set using the form's \texttt{action} property.
The body of the web request is a collection of name-value pairs encoded by the \texttt{FormSubmission} object.

\subsubsection{Browser \texttt{webRequest} API}
\label{sec:webrequest}

\begin{figure}
	\centering
	\includegraphics[width=6.5cm]{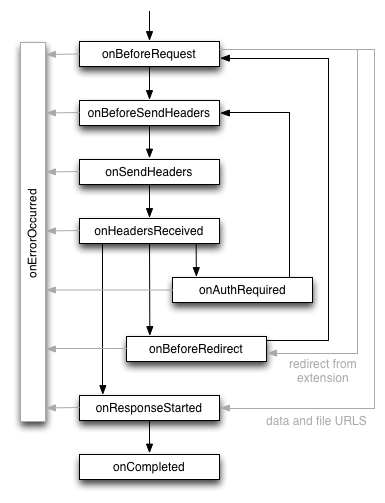}
	\caption{Web request API flow~\cite{chrome_webrequest}}
	\label{fig:webrequest}
	\Description{Shows the web request API flow in modern browsers.}
\end{figure}

The \texttt{webRequest} API is a browser extension-only API that lets extensions read, modify, or cancel web requests and responses.
To access this API, extensions register event listeners for one or more stages of the web request processing lifecycle (see Figure~\ref{fig:webrequest}).

In this lifecycle, the \texttt{onBeforeRequest}, \texttt{onBeforeSendHeaders}, and \texttt{onSendHeaders} events happen before the request is sent and provide access to the request body, allowing passwords included there to be exfiltrated.
The \texttt{onBeforeRequest} and \texttt{onBeforeSendHeaders} allow extensions to redirect or cancel the request, while the latter also allows modification of request headers.
The remaining stages focus on what happens after the request has been submitted. Since they lack access to the request body, they are less important in the context of secure password entry.

Critically, none of these stages allow the request body to be modified.
By the time \texttt{onBeforeRequest} is reached, the form submission process has already created the \texttt{FormSubmission} object, and the extensions are only provided with a read-only copy of the data in this object (not the object itself).
The reason behind not allowing the request body to be changed is unknown, but requests to add this functionality have gone unfilled over the last decade, and there is little evidence that it will be added~\cite{chromium91191webrequest, mozilla1376155webrequest}.
%Interestingly, functionality allowing the modification of request bodies existed in older browser versions (e.g., Firefox, Safari, Internet Explorer) but was removed when these browsers migrated the Chromium model for browser extensions.
This keeps the protocol that Stock and Johns described from working with modern browsers~\cite{stock2014protecting}.

%While the request body cannot be modified, the response body can be modified in the \texttt{onResponseStarted} stage.
%This allows extensions to arbitrarily inject client-side scripts into the webpage.
%However, this ability to modify response bodies requires additional permissions (\texttt{declarativeNetRequest}) on top of those required to read the body.

%% file: threats.tex
%!TEX root=paper.tex

\subsection{DOM-Centric Threat Model}
\label{sec:threat}

Stock and Johns' did not provide a threat model for their defense.
As such, we have created the following threat model for DOM-based password exfiltration.

\subsubsection{Attack Vector}
After a password is entered or autofilled into a webpage, that password is accessible as part of the DOM to JavaScript running on the page.
This JavaScript can be sourced from the webpage itself, JavaScript libraries loaded by the webpage, JavaScript injected by browser extensions, or JavaScript injected using an attack (e.g., cross-site scripting).

\subsubsection{Adversaries}
Based on the above two attack vectors, we have identified three local adversaries:

\paragraph{DOM Observer}
A DOM observer is an entity that the webpage itself has given access to the DOM to enable some desired functionality, such as serving ads or providing a JavaScript library.
As such, this adversary has full control of the webpage's DOM (attack vector 1) and can read and modify it.
While this entity does not intend to exfiltrate passwords, during normal operations, it may scan the DOM, unintentionally access the password, and send it back to a non-authentication-related endpoint~\cite{senol2022leaky}.
While this adversary may not be interested in the passwords themselves, other attackers can attempt to steal them from this entity if they store them (e.g., in their logs).
This entity is not intentionally malicious, so it will not attempt to circumvent password exfiltration defenses.

A good example of this type of adversary is web trackers.
Web trackers are scripts that track users across different websites to serve advertisements more effectively.
Trackers collect various information and user interactions without the user knowing about them.
Senol et al.~\cite{senol2022leaky} analyzed form submissions on 100,000 popular websites, investigating whether web trackers on those pages would exfiltrate user passwords.
They find that at least 2--3\% of those pages include web trackers that exfiltrate passwords.
While the motivations for this exfiltration are largely unknown, in many cases, it likely represents an honest-but-curious attacker model---i.e., web trackers are simply grabbing whatever information they can and are not targeting passwords specifically.
Regardless, as shown by Dambra et al.~\cite{dambra2022sally}, users encounter these web trackers frequently, indicating a need to protect user passwords against these honest-but-curious trackers.

\paragraph{Malicious DOM Exfiltrator}
This adversary has full control of the webpage's DOM (attack vector 1) and can read and modify it.
They have gained this access by inserting a malicious client-side script into the webpage.
They aim to leverage this access to retrieve the user's password from within the DOM.
They will succeed at their goal if the password is ever included in the webpage's DOM.
%They can also attempt to trick password managers into autofilling passwords outside the legitimate login page~\cite{silver2014password,stock2014protecting,oesch2020that}.
However, this adversary's capabilities are limited by security primitives built into the browser---they cannot violate the same origin policy (e.g., directly access passwords stored in the password manager) and do not have access to view the web requests created by the browser.

The most common source of malicious client-side scripts is cross-site scripting (XSS) attacks.
In these attacks, an adversary coerces a website to include attacker-controlled scripts in a web page's DOM, either by tricking users into clicking a link with the malicious script (reflected XSS attack) or uploading the malicious script to the website (stored XSS attack).
While defenses for XSS attacks are well known, the OWASP foundation consistently ranks them in its top 10 web application security risks~\cite{xssowasp}, and hundreds of XSS attacks have been reported in January of 2024 alone~\cite{CVEdetails2024}. 

Another common source of malicious client-side scripts are websites that use third-party libraries, which are ubiquitous~\cite{lemos2021dependency}.
While libraries produced by an adversary are transparently dangerous, more concerning are supply chain attacks~\cite{ohm2020backstabber,duan2020towards}.
In these attacks, an adversary compromises an otherwise benevolent software library, which then compromises websites that rely on this library when it is updated.
These attacks are already somewhat common~\cite{ohm2020backstabber,duan2020towards}, with WhiteSource~\cite{whitesource2022vulnerablenpm} identifying 1300 malicious JavaScript libraries in 2021.
Also, in 2024, an attack against a popular JavaScript library injected malicious code into over 100 thousand websites~\cite{sansec2024polyfill}.

While supply chain attacks are problems for client-side and server-side libraries, this paper is primarily concerned with client-side supply chain attacks.
Web client-side supply chain attacks are especially concerning as it is common to load libraries from external sources.\footnote{For example, see usage instructions for Bootstrap at \url{https://getbootstrap.com/}.}
If a website adopts this practice, it will be compromised immediately as soon as the library is compromised, without needing to wait for the website to explicitly update to the compromised version of the library.

%% file: sbc-pwm.tex
%!TEX root=paper.tex

\subsection{Nonce-Based Password Replacement}

To address DOM-based password exfiltration, Stock and Johns~\cite{stock2014protecting} proposed a nonce-based password replacement protocol.
In this protocol, password manager browser extensions autofill a random password (i.e., a nonce) instead of the actual password.
It is this random password that DOM-based scripts have access to, not the actual password.
The password manager also monitors all outgoing \texttt{POST} requests, looking for password nonces and replacing them with the actual password.

%While this paradigm is promising, it has two key limitations.
%First, it fails to protect against the \textit{Web Request Exfiltrator} adversary identified in our threat model, as this adversary simply reads the password from the final POST request.
%Second, its design is incompatible with modern browser design and can not be implemented.

\section{A Secure Browsers Channel for Password Autofill}
\label{sec:sbc-pwm}

Unfortunately, even at the time Stock and Johns' proposed their nonce-based password replacement protocol, it was incompatible with modern browser design~\cite{chromium91191webrequest}, and the one browser where it did work, Firefox, removed this functionality shortly after~\cite{mozilla1376155webrequest}.
Moreover, there was no obvious path for getting this protocol to work in browsers, short of the daunting task of figuring out how to modify the browsers themselves.
This reality prevented this protocol from being able to be adopted by password managers~\cite{oesch2020that}.
Additionally, it kept researchers from being able to build on this idea.

Our paper's first contribution is addressing these challenges by describing how browsers can be modified to provide a nonce-based password replacement API.
Implementing this API was non-trivial, requiring hundreds of hours of research and development work.
Not only does our contribution provide a direct benefit to password manager users, but it also benefits the research community by providing a working implementation that they can build on to continue exploring what is possible with nonce-based replacement defense---something that was not previously possible, but for which there are many areas of compelling future work (see \S\ref{sec:fw}).

%To address these limitations, we propose an alternative design, replacing the extension-based secure channel proposed by Stock and Johns' with a browser-based secure channel of our design.
%Our design addresses the above limitations, fully protecting against local attacks against the password autofill workflow.
%We also implement our design, empirically demonstrating its feasibility, security, and deployability.
%To promote open science, this implementation can be found in the Firefox repository (branch names redacted for anonymity).

\subsection{API Design}
\label{sec:api:design}
Our API is an extension of the Web Request API found in all modern browsers, adding a new stage to the \texttt{webRequest} API that supports nonce-based password replacement: \texttt{onRequestCredentials}.
This stage occurs before \texttt{onBeforeRequest}.
This new stage provides a secure browser channel for transmitting passwords between password managers and websites without allowing any DOM-based scripts access to the content.

To use this channel, a password manager will first autofill a random password nonce into the webpage.
At the same time, it will register a callback handler with \texttt{onRequestCredentials}.
This handler will be called during page submissions, receiving read-only details about the web request.
Using these details, the password manager will determine if replacing the password nonce with the actual password is safe.
Based on recommendations from the research literature~\cite{stock2014protecting,li2014emperor,silver2014password,oesch2020that,oesch2021emperors} and to prevent a reflection attack discussed in \S\ref{sec:reflection}, our design requires password managers to make the following safety checks:

\begin{enumerate}
	\item Check that the web page is not displayed in an \texttt{iFrame}.
	
	\item Check that the submission channel does not use HTTP or an insecure HTTPS connection.
	
	\item Check that the origin (protocol, domain, port) matches the origin identified by the password manager entry being autofilled. Even better, the password manager can store the exact URL where credentials should be submitted, only submitting passwords to that URL.\footnote{Determining the exact URL could be done by having password managers store associations for popular websites, storing prior successful submission locations, or crowdsourcing the creation of associations.}
	
	\item Check that the nonce is not in the GET parameters.
	
	\item Check that the name of the field storing the nonce matches the name of the field where the password nonce was autofilled. Additionally, this field should have a valid name, such as \texttt{password}~\cite{oesch2020that,oesch2021emperors}.
\end{enumerate}

If the handler determines a substitution should take place, it returns a tuple with the following data:
\begin{enumerate*}[label=\emph{(\roman*)}, itemjoin={{, }}, ]
	\item the name of the field that was autofilled with the password nonce
	\item the password nonce
	\item the replacement password
	\item the origin (scheme, host, port) associated with the password
\end{enumerate*}.
After receiving this data from registered handlers, the browser will then search through the key-value pairs in the \texttt{FormSubmission} object (see \S\ref{sec:formsubmission}), making the requested substitution if and only if
\begin{enumerate*}[label=\emph{(\alph*)}, itemjoin={{, }}, ]
	\item the indicated field's value matches the password nonce exactly
	\item the web request will be sent to the indicated origin
\end{enumerate*}.
While the password manager should already have checked these details, we have the browser also check them to provide some level of defense in depth.
Because replacement happens within the \texttt{FormSubmission} object, the browser handles proper sanitization and encoding of the key-value pairs.
From this point on, the browser processes the web request as normal.

\begin{figure*}
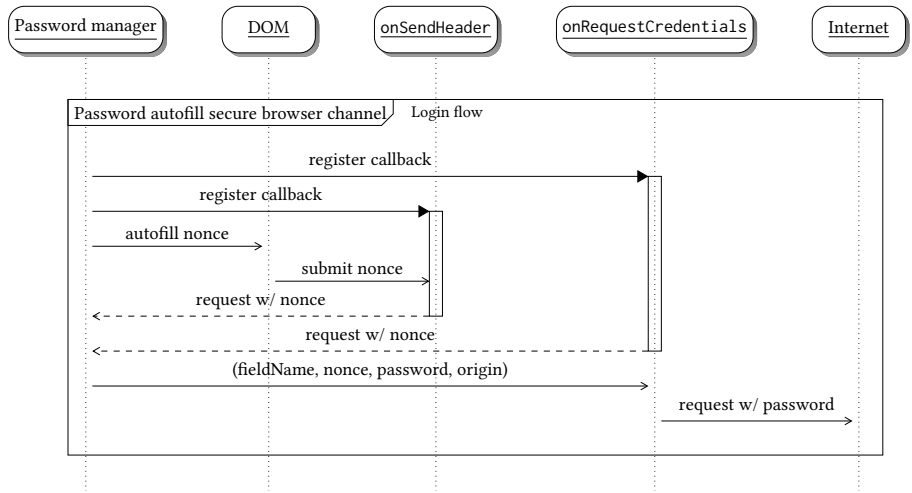

	\centering
	\adjustbox{max width=.7\textwidth}{
		\begin{sequencediagram}
			\tikzset{inststyle/.append style={
					drop shadow={top color=gray, bottom color=white}, 
					rounded corners=2.0ex
				}
			}   
			
			\newinst{p}{Password manager}
			\newinst[1]{d}{DOM}
			
			\newinst[1]{b}{\texttt{onSendHeader}}
			\newinst[1]{a}{\texttt{onRequestCredentials}}
			\newinst[1]{i}{Internet}
			
			\begin {sdblock}{Password autofill secure browser channel}{Login flow}
			\begin{call}{p}{register callback}{a}{request w/ nonce}
				\begin{call}{p}{register callback}{b}{request w/ nonce}
					\mess{p}{autofill nonce}{d}
					\mess{d}{submit nonce}{b}
					
				\end{call}
			\end{call}
			\mess{p}{{(}fieldName, nonce, password, origin{)}}{a}
			\mess{a}{request w/ password}{i}
		\end{sdblock}
	\end{sequencediagram}
}

\caption{A flow diagram for our password autofill secure browser channel}
\label{fig:design}
\Description{Gives the workflow for the proposed defense. This image demonstrates that only the nonce, not the password is ever available to the DOM or to malicious browser extensions.}
\end{figure*}

The above flow is visualized in Figure~\ref{fig:design}.
Looking at this flow, it is clear that the password nonce is ever available within the DOM, preventing password theft by either the \textit{DOM Observer} or the \textit{Malicious DOM Exfiltrator}.

\subsection{API Implementation}
\label{sec:implementation}

%\begin{figure*}
%	\centering
%	\adjustbox{max width=.7\textwidth}{
%		\begin{sequencediagram}
	%			\tikzset{inststyle/.append style={
			%					drop shadow={top color=gray, bottom color=white}, 
			%					rounded corners=2.0ex
			%				}
		%			}   
	%			
	%			\newinst{p}{Password manager}
	%			\newinst[1]{d}{DOM}
	%			
	%			\newinst[1]{b}{\texttt{onBeforeRequest}}
	%			\newinst[1]{a}{\texttt{onRequestCredentials}}
	%			\newinst[1]{i}{Internet}
	%			
	%			\begin {sdblock}{Browser-Based Nonce Injection}{Login flow}
	%			\begin{call}{p}{register callback}{a}{request details w/o the request body}
		%				\begin{call}{p}{register callback}{b}{request details w/ nonce}
			%					\mess{p}{autofill nonce}{d}
			%					\mess{d}{submit nonce}{b}
			%					
			%				\end{call}
		%			\end{call}
	%			\mess{p}{[fieldName: <nonce, password>]}{a}
	%			\mess{a}{request w/ password}{i}
	%		\end{sdblock}
%	\end{sequencediagram}
%}
%
%\caption{A flow diagram for our password autofill secure browser channel}
%\label{fig:implementation}
%\end{figure*}

To validate our design's feasibility, security, and deployability, we implemented it into Mozilla Firefox 107.0 and the Bitwarden password manager.
Below, we describe the implementation details for injecting and replacing password nonces.
%Appendix~\ref{appx:porting} discusses how our implementation could be ported to other browsers.

\subsubsection{Step 1: Injecting the Nonce}
First, the password manager will register two callbacks, one for \texttt{onBeforeRequest} and one for the new \texttt{onRequestCredential}.
Next, as needed, the password manager will autofill a nonce in place of a password.
After doing so, it will internally store an association between the web page and 
\begin{enumerate*}[label=(\roman*),itemjoin={,\space{}},itemjoin={,\space{}and\space{}}]
\item the nonce
\item the name of the field storing the nonce
\item the password manager entry that is being autofilled 
\end{enumerate*}

\subsubsection{Step 2: Validating the Replacement} \label{sec:validating}
Eventually, the page will be submitted, and a \texttt{webRequest} will be created by the browser, with this \texttt{webRequest} containing the autofilled nonce.
This will cause the callback registered with \texttt{onBeforeRequest} to be triggered.
Within this callback, the password manager can see the details of the \texttt{webRequest}, including the destination URL, the HTTP method used, and the request body.
Using this information, the password manager determines if (a) there is a nonce associated with the web page that generated the \texttt{webRequest}, (b) the nonce is in the request body, and (c) replacing the nonce would be safe.
If all these checks pass, the password manager will internally store an association between the \texttt{webRequest} and 
\begin{enumerate*}[label=(\roman*),itemjoin={,\space{}},itemjoin={,\space{}and\space{}}]
\item the nonce
\item the name of the field storing the nonce
\item the password manager entry that is being autofilled 
\end{enumerate*}

\subsubsection{Step 3: Executing the Replacement}
Immediately before sending the \texttt{webRequest} over the wire, the callback registered with \texttt{onRequestCredential} will be triggered.
This callback once again receives request details as input, with the request body stripped out.
%This is necessary to prevent extensions from reading any nonce substitutions that might have been made by other \texttt{onRequestCredential} callbacks.
In this step, if the password manager has a nonce associated with the current \texttt{webRequest}, it will simply return the associated nonce, password, field storing the nonce, and the URL to which the password manager expects the password to be submitted.

The password manager will then use this information to locate and replace the nonce in the request body.
To prevent unintended (or malicious) consequences from replacing the nonce, the browser will only make this substitution if,
\begin{enumerate}[label=\roman*)]
\item The field name storing the nonce exactly matches the field name specified by the password manager.
\item The field value exactly matches the nonce with no additional characters.
\item The submission URL matches the origin (protocol, domain, port) of the URL provided by the password manager.
\end{enumerate}

While the password manager should have already checked these items before submitting the password to be replaced, we have the browser also check these items to ensure they will be checked.
This is an important form of defense in depth~\cite{oesch2021emperors}.
If any of these checks fail, no substitution is made, and the request will be sent still containing the nonce.

%% file: threats-full.tex
%!TEX root=paper.tex

\section{Handling Malicious Extensions}

As we designed our API, we identified a limitation in the threat model used by Stock and Johns' protocol---it did not consider malicious browser extensions.
While such extensions were rare when Stock and John's paper was published, they have become a significant problem in our modern day~\cite{kapravelos2014hulk,pantelaios2020you,kaspersky2023dangerous}.
As such, our paper's second contribution is providing an extension to the base threat model to account for malicious extensions.
We also describe how our API was designed to protect against this class of attacker.

\subsection{Malicious Extension Threat Model}
\label{sec:threat-extended}

\subsubsection{Attack Vectors}

In addition to exfiltrating the password from the DOM, a malicious extension can also try to exfiltrate the password from an outgoing web request.
When the form containing the password is submitted, the browser creates a web request to transmit the password to the server.
Browser extensions with permissions for the \texttt{webRequest} API can read the content of outgoing web requests, exfiltrating it there.

\subsubsection{Adversary---Web Request Exfiltrator}

This adversary can view web requests and their associated data before the browser transmits them.
They have gained this access by tricking the user into installing a malicious browser extension or compromising an extension the user has already installed.
Their goal is to exfiltrate passwords from the web request's body just before transmission.
This attacker may also be able to inject arbitrary JavaScript into the webpage.
However, as this capability is already covered by the malicious DOM exfiltrator, for this attacker, we focus primarily on their ability to read web requests.

There are many instances of malicious browser extensions used by millions of users on official Chrome/Firefox extension stores.
In 2017, eight extensions, used collectively by over 4.8 million users, were compromised and used to steal login credentials~\cite{munder2017psa}.
In 2020, 500 Chrome browser extensions were discovered secretly uploading private browsing data to attacker-controlled servers and redirecting victims to malware-laced websites~\cite{ositcom}.
Also, Awake has identified 111 malicious Chrome extensions that take screenshots, read the clipboard, harvest password tokens stored in cookies or parameters, grab user keystrokes (like passwords), etc.~\cite{awakesecurity}. 
These extensions have been downloaded over 32 million times.
Finally, in 2021, Cato's analysis of network data showed that 87 out of 551 unique Chrome extensions used on customer networks were malicious~\cite{catonetworks}.
\emph{In addition to inherently malicious extensions, extensions can be compromised through supply chain attacks, as already discussed for the DOM attacker.}

There are limitations on what a malicious browser extension can do.
Most importantly, all extensions are given a unique origin, preventing one extension from accessing the data or scripts of another extension.
As such, the browser's same-origin policy prevents malicious extensions from directly accessing each other's data or scripts.
The only way for a malicious extension to gain access to the passwords stored by a password manager is for that manager to copy those passwords to an origin where the malicious extension has access (i.e., the webpage).

The second limitation is that for a malicious extension to exfiltrate passwords, it needs to be granted one or more of the following permissions~\cite{chromeManifest}:

\begin{itemize}
	\item The \texttt{scripting} permission allows the injection of client-side scripts on any webpage~\cite{chromeScriptingAPI}.
	\item The \texttt{activeTab} permission allows the injection of client-side scripts on the root webpage of the actively focused tab~\cite{chromeActiveTab}. 
	\item The \texttt{content script} attribute injects a specified client-side script into webpages with a matching origin (wildcards are allowed)~\cite{chromeContentScripts}.
	\item The \texttt{declarativeNetRequest} permission allows the modification of web response bodies, which can be used to inject client-side scripts~\cite{chromeDelaritiveNetRequestAPI}.
	\item The \texttt{webRequest} permission allows reading web request bodies (which can contain passwords)~\cite{chromeWebRequestAPI}.
\end{itemize}

For an extension to be granted these permissions, the user must approve the grant during installation or when the permission is first used.
However, research has shown that users struggle to understand these permissions and are likely to grant them without much consideration~\cite{korir2022empirical}.
Still, permissions are useful---if an extension is compromised through a supply chain attack, the attacker will not be able to change the manifest and will be constrained by the extension's existing permissions.

To understand the prevalence of these permissions in existing extensions, we collected 101,414 browser extensions from the Chrome webstore between August 28, 2022, and September 13, 2022.
We then extracted and analyzed their manifest files.
We also estimate user counts based on data fetched from CRXcavator\footnote{\url{https://crxcavator.io/}} (pulled May 03, 2023).

We identified 12,576 extensions that can inject client-side scripts on any webpage, including 55 with at least 500,000 users.
We also identified 4,169 extensions that can read any web request's body, including 19 with at least 500,000 users.
Of these extensions, 1,410 can read any web request's body but not inject client-side scripts, including 6 with at least 500,000 users.
While it is unclear how many of these extensions are malicious (the percentage is likely small), it is not unheard of, even for extensions with many downloads~\cite{kaspersky2023dangerous}.
Moreover, these extensions are vulnerable to compromise through supply-chain attacks, which are becoming increasingly common.

\subsection{API Design}

In the original design of our API, malicious extensions can access the password as it will be visible in the \texttt{requestBody} after it is replaced in \texttt{onRequestCredentials}.
To address this flaw, we make two changes to our API.
First, we move onRequestCredentials after all stages where \texttt{requestBody} is accessible: after \texttt{onSendHeaders} and before \texttt{onHeadersReceived}.
Second, when presenting extensions with the \texttt{requestBody} in \texttt{onRequestCredentials}, we show \texttt{requestBody} as it was before any substitutions were processed.
This prevents malicious extensions from learning about the password manager's replacement value (i.e., the password) during \texttt{onRequestCredentials}.
Moreover, since this is the last stage where \texttt{requestBody} is accessible using the \texttt{webRequest} API, they will not have an opportunity to view the modified \texttt{requestBody}.

\subsection{API Implementation}
In our implementation, we slightly deviated from the design described above.
Instead of having the web request's body available in the \texttt{onRequestCredentials} callback, we stripped this data entirely from callbacks at this stage.
This necessitated moving the password manager's code, validating whether a replacement should occur into the \texttt{onBeforeRequest} event, which has access to the request body.
Replacement still occurs within the \texttt{onRequestCredentials} stage, meaning that our implementation still protects against password exfiltration from browser extensions.
%Figure~\ref{fig:implementation} shows the workflow as implemented.

This deviation arose from an implementational challenge around creating copies of the request body, which is necessary to prevent malicious extensions from seeing replaced passwords during the \texttt{onRequestCredentials} stage.
As this only further restricts information available to extensions, it does not negatively impact security.
We eventually resolved the challenge that prevented us from implementing our solution exactly as described in our design; however, by then, we had already conducted a thorough evaluation of its functionality and performance, so we described the implementation as it existed when conducting this testing.

%% file: sbc-pwm-evaluation.tex
%!TEX root=paper.tex

\section{Evaluation}

\subsection{Security} \label{sec:security-evaluation}

By design, our API prevents DOM-based and \texttt{webRequest}-based exfiltration.
The password manager never injects the nonce into the DOM, and JavaScript running within a web page has no access to the \texttt{webRequest} API.
As such, there is no way for DOM-based adversaries to access the secret.
Similarly, as \texttt{onRequestCredential} is the final stage where the \texttt{requestBody} is available, and it does not share the modified \texttt{requestBody} with extensions, malicious extensions are unable to access the substituted password.

Still, to provide even more confidence in our defense, we also tried to conduct DOM-based attacks to see if there were any mistakes in our API or password manager's implementation.
To this end, we created a basic PHP login page that takes in a username and password as form input and logs the credentials in plaintext for verification purposes.
To test this web page, we would autofill with the password and then submit the web page.
To simulate a DOM-based attack, we would inject arbitrary JavaScript on the test web page, attempting to scrape the password field before page submission.
To simulate extension-based attacks, we built extensions with full access to the \texttt{webRequest} API, including the new \texttt{onReplaceCredential} callback.

As expected, these efforts yielded no fruit.
Still, they were valuable as they helped us identify a potential attack that can occur if the password manager does not check the destination URL where the password will be sent---a \textit{reflection attack}.
This attack is prevented by following the safety checks described in \S\ref{sec:api:design}.

%Besides this attack, we could not identify any JavaScript or extension calls that gave us access to the replaced credential.

\subsubsection{Reflection Attack}
\label{sec:reflection}

An attacker aware that password nonces are being used could attempt a reflection attack.
In this attack, they would do the following:

\begin{enumerate}
	\item Change the password form to submit to a page that displays (i.e., reflects) one or more of the submitted form fields. This page needs to be within the same origin to trigger replacement.
	\item Change the name of the password field to one of the fields that will be reflected.
	\item Cause the form to be submitted.
	\item Read the reflected password from the vulnerable page. The attacker will need JavaScript access to that web page to do this.
\end{enumerate}

An example of this attack is changing the field name for the password to ``username'' for web pages where an error is displayed when the username does not exist (e.g., ``\$\{username\} does not exist'').

For the most part, our implementation is not vulnerable to this attack.
First, autofill will often fail if the attacker changes the password field's name before autofill.
Second, if the attacker changes the password field's name after autofill, the nonce will not be replaced, as our implementation only replaces the nonce when it is in a field named the same as the field where the nonce was injected.

Still, this attack may work in cases where the attacker finds a webpage where all fields are reflected.
To address this, password managers can store the URL where passwords are submitted and only replace the password when sent to that destination.
This list of URLs can either be pre-generated by the password manager provider for popular websites or stored the first time the password is autofilled and transmitted by the browser (a trust-on-first-use model).
This defense can be further strengthened by having the password manager learn the name of the password field, ensuring fields with different names are not autofilled.
As several password managers are already checking these items before autofilling passwords~\cite{oesch2020that}, we believe implementing these checks to be quite practical and have included them in our list of password manager safety checks described with our design.

\subsection{Deployability}
\label{sec:functional-evaluation}

To evaluate the functionality of our browser-based nonce injection implementation, we conducted tests on real websites to ensure that our implementation does not break their authentication flows.
%We implemented our measurement setup on a virtual machine with the Xubuntu operating system version 22.04, 16 GB of memory, and four cores on an i9 processor.
%
Using the Alexa top 1000 sites from May 1, 2022, we ran a Selenium script that started at the root page of each domain and traversed all links on the page up to a maximum depth of three,\footnote{We selected a depth of three pages empirically by manually analyzing the top 15 domains.} searching for login pages.
In total, we identified login pages on 623 sites.
%The script searched for password input fields in the pages and saved the links. 
After filtering for the subdomains of the same website, like Google and Amazon, for different countries, we were left with 573 unique login pages.

%TODO: This is an easy place to save space. Remove the figure.

\begin{figure}
	\includegraphics[width=\linewidth]{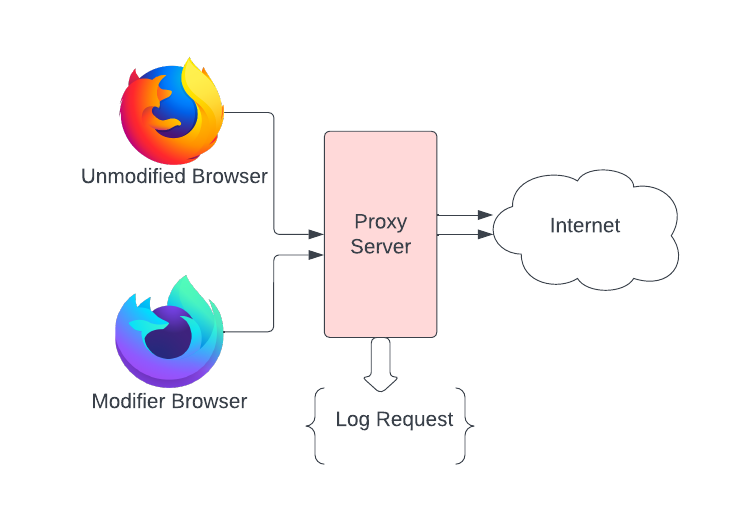}
	\caption{Functional Evaluation Architecture}
	\label{fig:testing_arch}
	\Description{Describes the setup for the functional evaluation of our prototype. Login events happen in both an unmodified and modified version of Firefox, with authentication flows captured by a proxy server. This server records the password transmitted over the network to determine if the nonce was properly replaced by our prototype.}
\end{figure}

For all 573 login pages, we used a Selenium script to submit credentials from an unmodified Firefox browser and our modified version of Firefox.
As illustrated in Figure~\ref{fig:testing_arch}, we set up a proxy server to record all the outgoing web requests from the browsers and save the request body. 
We then compared the credentials in the authentication requests sent from both browsers.

For 97\% of websites (554/573), the \texttt{webRequests} from both browsers was identical, indicating our defense would work without any modification to the website.
Of the remaining websites, 2\% (11/573) calculate a hash of the password in the outgoing \texttt{webRequest}.
Because the hash value is calculated by JavaScript running in the DOM, when using our defense, the calculated hash will be based on the nonce.
As such, any server-side integrity checks will fail and prevent the login attempt.
%Due to a lack of accounts at these websites, we couldn't check if such a check is made server-side, though it likely is.
Similarly, 1\% of websites (8/573) modify the password (e.g., base64 encoded, hashed) before submitting it.
This prevents our defense from detecting the nonce in the outgoing \texttt{webRequest}, meaning that the password will not be sent over the wire.
On the websites where our defense is incompatible, the login will fail, and users will be required to manually enter the password into the login page, bypassing our defense.
Overall, our measured incompatibility rate is similar to the rates identified by Stock and Johns'~\cite{stock2014protecting} and by Li and Evans~\cite{li2017horcrux}.

%The credentials submitted to the websites were the same in 554 of 573 cases (97\%).
%Eleven sites (2\%) generated an integrity check value based on the autofilled nonce.
%In these cases, the nonce would still be replaced by the real password, but the integrity check value would no longer match.
%While we did not confirm that this would cause the login to fail (we did not have actual accounts on these sites), we still consider these as failed cases.
%Finally, seven websites (1\%) either hashed or base-64 encoded the nonce, which prevented the password manager from replacing it with the real password.

While the incompatibility rate of our defense is similar to the failure rate of password autofill in general~\cite{huaman2021case}, this rate can still be a usability annoyance~\cite{huaman2021case,oesch2022observational}.
This annoyance could drive some users away from password autofill and could make password managers hesitant to implement our defense.
In the short-term, we recommend that password managers implementing our defense maintain a list of non-compatible websites for which they will not inject password nonces, obviating this problem.
As our testing shows, identifying non-compliant websites can be fully automated, making creating such a list highly feasible.
In the long-term, future work can bridge the gap towards 100\% compatibility (see~\S\ref{sec:fw:compatibility}).

\subsection{Overhead}
\label{sec:overhead-evaluation}

To measure the overhead incurred by our implementation, we examined the logs for \texttt{HttpBaseChannel} generated during our functional evaluation of our modified browser.
These logs give exact measurements for each stage of the \texttt{webReuqest} lifecycle, allowing us to pinpoint the processing time used by our code.

When a replacement is needed, we measured the average \texttt{webRequest} lifetime to be 4.5222 seconds.
Our code accounted for 0.443 seconds, representing 10.6\% of the total request duration.
While this is already a minimal overhead, we also note that authentication is an uncommon event, meaning that users are unlikely to experience this overhead more than a few times a day.

This overhead can entirely be explained by the time needed to destroy and recreate the web request's body when replacing the nonce with the passwords.
Based on our experience with the browser's internals and the \texttt{webRequest} API, there is no way to avoid this overhead without entirely re-engineering the API to change when it serializes request bodies.
While this is certainly possible, our implementation tried to avoid making large-scale changes to make it more palatable for browser developers to adopt.

We also measured the overhead of our API changes when a replacement was unnecessary, finding no impact on performance (i.e., it was indistinguishable from noise).

%% file: porting.tex
%!TEX root=paper.tex

\newcommand*{\code}[1]{{\small \texttt{#1}}}

\section{Implementing in Other Browsers}
\label{appx:porting}

The three major browser engines, Chromium (Chrome, Edge, Opera, etc.), WebKit (Safari), and Gecko (Firefox), account for 96\footnote{https://gs.statcounter.com/browser-market-share}--99\%\footnote{https://www.statista.com/statistics/545520/} of browser market share.
While the codebases for each engine are distinct, porting our defense from Gecko to Chromium and WebKit is simplified by the fact that all three browser engines implement the standardized webRequest API, which forms the backbone of our defense.

\subsection{Porting using the \code{webRequest} API}

To port our defense to other browsers, it is necessary to implement the following functionality:

\begin{enumerate}
	\item \textbf{Gather nonce replacements:} Add a \code{onRequestCredential} event to the \code{webRequest} API, allowing the browsers to gather nonce replacements from extensions.
	\item \textbf{Execute nonce replacements:} Modifying the networking code to fire the \code{onRequestCredentials} event and modify the \code{webRequest} based on received nonce replacements.
\end{enumerate}

\subsubsection{Chromium}

\paragraph{Gather nonce replacements}
Communication with the extensions will occur in a new \code{WebRequestEventRouter:: OnRequestCredentials} method. % (\code{extension\_web\_request\_event\_router.cc}).
The existing \code{OnBeforeRequest} method can be used as an example of how to implement this method.
Within \code{OnRequestCredentials}, extensions with registered listeners will be called in a blocking fashion, each returning their list of requested nonce replacements (if any).
After validating these nonce replacements (see \S\ref{sec:api:design}), each will be stored in \code{WebRequestEventRouter::response\_deltas} as an \code{EventResponseDelta} %(\code{web\_request\_api\_helpers.cc}), 
with this class being updated to support storing nonce substitutions.

To allow extensions to register a listener for \code{OnRequestCredentials} to call, this event will be added to the \code{web\_request.json} file.

\paragraph{Execute nonce replacements}
The \code{WebTransportHandshakeProxy:: OnBeforeRequestCompleted} will be modified to replace its call to \code{OnBeforeSendHeaders} with a call to \code{OnRequestCredentials}.
The completion handler for \code{OnRequestCredentials} will be a new \code{OnRequestCredentialsCompleted} method.
This new method will be responsible for calling the previously replaced \code{OnBeforeSendHeaders} method (and its respective \code{OnBeforeSendHeadersCompleted} method).

At the bottom of the \code{OnRequestCredentialsCompleted} function, the \code{ExecuteDeltas} function will handle the modification of the body.
\code{ExecuteDeltas} will need to be modified to support nonce replacements.

\subsubsection{WebKit}

\paragraph{Gather nonce replacements}
Communications with the extensions will happen in the existing \code{NetworkResourceLoader::startNetworkLoad} method, after the \code{ResourceLoadDidSendRequest} is sent to extensions.
Communication involves an inter-process communication (IPC) message being sent to the web page's process.
This can be implemented by adding a message receiver in \code{NetworkProcessProxy.messages.in} and defining the code for that receiver in the \code{NetworkProcessProxy} class.
This message will then be forwarded to \code{WebPageProxy}, followed by \code{WebExtensionController}, then \code{WebExtensionContext}.
A second IPC message will be sent to the web extension's process, which must register to receive it in its \code{WebExtensionAPIWebRequest.messages.in} file and register a handler in \code{WebExtensionAPIWebRequest}.
Finally, this method can call registered listeners, each of which returns its list of requested nonce replacements (if any).
Valid nonce replacements (see \S\ref{sec:api:design}) will be returned through this IPC call chain.
This chain of passed messages can be modeled off a similar messaging chain started for \code{resourceLoadDidSendRequest} in \code{NetworkResourceLoader:: startNetworkLoad}, except that the \code{send()} method needs to be replaced with the \code{sendSync()} method, allowing responses to be passed back using IPC.

To allow extensions to register a listener for the \code{onRequestCredentials} event, it will need to be added to \code{WebExtensionAPIWebRequest.idl}, with code to create the event added in \code{WebExtensionAPIWebRequest}.

%Note, all extension-specific functionality is implemented using Cocoa, meaning that when searching for class files, the file inside the cocoa folder should be used in place of the default class implementation.

\paragraph{Execute nonce replacements}
Immediately after getting the list of nonce replacements, the web request's body can be directly modified within \code{startNetworkLoad}.
This is straightforward and can follow the approach taken in our PoC Firefox implementation.

\subsection{Porting For Manifest V3}
\label{appx:porting:manifestv3}

Our password defense is implemented based on Manifest v2, which is supported by all three major browser engines.
However, it is possible to modify the implementation of our defense to support Manifest V3.
The major difference for Manifest v3 is that it is no longer possible to query extensions for responses during the webRequest workflow.
This necessitates finding another mechanism for providing information on requested nonces to the code modifying the web request (which is otherwise identical to the Manifest v2 implementation).

To do this, a new API is added to the window (e.g., \code{window.secrets}), with a \code{registerNonce} method that takes as input a nonce, a secret value (i.e., the password), and a policy regarding when the nonce should be replaced by the password in outgoing requests (see \S\ref{sec:api:design}).
This method will store these values within the \code{Page} object.
To implement this method, (i) a new \code{idl} file will be created to define the API, (ii) the API will be annotated to to require the \code{secrets} permission (or other appropriately named permission), and (iii) the \code{registerNonce} method will be annotated to receive the script state, giving it access to the \code{Page} object, (iv) a header and class file for the API will be written implementing the \code{registerNonce} functionality.
Adding the IDL's is straightforward, as they are all based on the Web IDL standard.
Implementing the new functionality is also straightforward, as existing IDLs provide a clear guide on how to interact with the \code{Page} object and can be used as a model for adding this functionality.

Next, code generating web requests (e.g.,, \code{FormSubmission::populateFrameLoadRequest},  \code{XMLHttpRequest:: createRequest}, \code{FetchRequest::create}) will be modified to retrieve the nonce substitutions from the \code{Page} object and include them with the web request details.
When the code executes, the nonce replacement is run, which retrieves the nonce substitutions from the web request information.
It will be responsible for validating the replacement using its internal logic (see \S\ref{sec:api:design}) and the policy given when the nonce replacement was registered.
%As the contents of the browser engine's internal web request object are not accessible to the DOM or extensions (verified by the authors), the stored passwords remain safe from exfiltration.

In comparison to our initial defense, the drawback of the Manifest V3 approach (and why we choose not to implement our defense in this fashion) is that it prevents password managers from performing policy checks on the nonce replacement; thus, the browser becomes a single point of failure for policy checks.
While this is not necessarily a problem, having two entities checking the nonce replacement can improve security~\cite{oesch2021emperors}.

%\paragraph{Chromium}
%
%\paragraph{WebKit}
%The new API will be added to the \code{WebCore/page} directory.
%Code for adding nonce replacements to requests will need to be added to \code{FormSubmission::populateFrameLoadRequest}, \code{XMLHttpRequest::createRequest}, and \code{FetchRequest::create}.
%The \texttt{ResourceRequestBase} is the class that will store the nonces at this point.

%% file: sbc-2fa.tex
\newcommand*{\rp}{\texttt{RP}\xspace}
\newcommand*{\rps}{\texttt{RP}s\xspace}
\newcommand*{\hsk}{\texttt{HSK}\xspace}
\newcommand*{\hsks}{\texttt{HSK}s\xspace}

\section{A Secure Browser Channel for FIDO2}

As we built, designed, and implemented our API, we hypothesized that similar APIs could be used to harden other browser-server services.
Of particular interest to us was the FIDO2 protocol.
Recent work has shown that this protocol is vulnerable to DOM-based attacks and malicious extensions~\cite{guan2022formal,hu2016security,yadav2024security,mahdad2024breaching}, the same adversaries our API was able to thwart in terms of password exfiltration.

As such, the third contribution of our paper is the design and implementation of a FIDO2 API, inspired by our nonce-based password replacement API, that protects FIDO2 against DOM- and malicious extension-based attacks.
This demonstrates that our approach generalizes to secure other browser-server interactions.
Moreover, it highlights the research importance of our first contribution: by providing a working implementation of nonce-based replacement for the browser, we have enabled research to build upon Stock and Johns' insightful protocol, where it has been otherwise stopped for the last 11 years. 

\subsection{FIDO2 Background}

FIDO2 is a web authentication protocol that leverages public-key cryptography.
A website stores the user's public key, and the corresponding private key remains with the user to sign challenges from the website during authentication.
There are three participants in FIDO2 authentication: 

\begin{itemize} 
	% [topsep=1pt,noitemsep,leftmargin=1.5em]
	\item An \textbf{authenticator} is a hardware device that holds the user's private key.
	FIDO2 supports a variety of client-side authenticators, including external (remote) authenticators like hardware security keys (\hsks) and built-in platform authenticators such as biometrics.
	Users provide input to an authenticator, such as a PIN or button press, to authorize it to login to a website. 
	
	\item A \textbf{website} is a web application like \texttt{github.com} that offers FIDO2 as a method for 2FA or passwordless authentication. 
	The website communicates with the authenticator through a WebAuthn client.

	\item A \textbf{WebAuthn client}, typically a web browser, acts as a relay between the website and the authenticator.
	To prevent phishing, the client provides the authenticator with the website's URL.
	The authenticator uses the credentials bound to the website to generate a response.
	The WebAuthn API in the client provides an interface to the website.

\end{itemize}

%\subsubsection{FIDO2 Registration and Authentication}
%\begin{figure*}
%	\centering
%	\includegraphics[width=.7\textwidth]{figs/FIDO2_registration.png}
%	\caption{FIDO2 registration protocol}
%	\label{fig:FIDO2_registration}
%\end{figure*}

Users register their \hsk as a method for 2FA or passwordless authentication. 
When a user clicks the Register button, the client sends a registration request to the website. Registration proceeds as follows:
(1) The website first sends a registration request to the webAuthn client. This request provides the webAuthn client with a challenge, user information, and website information. (2) The webAuthn client sends this information to the \hsk, including the website's URL and request type. (3) The \hsk asks for consent from the user, typically a key tap, and then generates a new asymmetric key pair. (4) The \hsk sends the credential id, public key, the website ID hash --- a SHA-256 hash of RP domain or sub-domain, counter --- to keep track of the count of authentication, and attestation signature: signed data object containing statements about the public key credential and the \hsk make and model to the webAuthn client as an attestation object. (5) The webAuthn client then forwards this data to the website. Once the website receives the data, it verifies the signatures and critical parts of the response and, if successful, registers the \hsk to the user's account.

Authentication mirrors registration, except (1) authentication omits the user information, and (2) attestation is replaced by the \hsk signing the response with the private key generated during registration.

%\subsubsection{FIDO2 Transaction Confirmation}
%FIDO2 Transaction Confirmation provides the relying party with the capability to ascertain not only the user's engagement in a transaction but also to verify the congruence of the transaction with the user's intended actions. This validation mechanism is realized through the transmission of transaction-specific information, as perceived by the relying party, to the user. Subsequently, this information is displayed to the user before the user approves the transaction.
%
%There are various FIDO2 extensions that allow transaction confirmation for various purposes. For example, Secure Payment Confirmation~\cite{spc} allows users to verify and authenticate payment transactions. No FIDO2 extension explicitly exists to allow users to verify registration and authentication requests. Simple Transaction Authorization Extension (txAuthSimple) allows \rp to enter any text shown to the user on the \hsk or browser. 

\subsection{Threat Model}

The threat model used in this section is based on the threat model for local attacks against FIDO2 described by Yadav and Seamons~\cite{yadav2024security}.
Attackers in this threat model have one of two goals: a \textit{short-term} goal where they trick the user into validating a single login request of the attacker's choosing and a \textit{long-term} goal where the attacker registers their authenticator in place of the user's authenticator for some account.
Both attacks can be achieved if the attacker can access and modify FIDO2 values sent from the server to the authenticator or from the web page to the server.
We identify two locations where a local attack can attack the FIDO2 protocol.

\begin{enumerate}
	\item \textbf{Modify values passed through the DOM.}
	An attacker can compromise the FIDO2 protocol by injecting malicious JavaScript into the webpage, executing the protocol, and redirecting the FIDO2 request/response to an attacker-controlled authenticator.
	One way to achieve this is by overriding the built-in \texttt{credentials.create} and \texttt{credential.get} calls from the \texttt{webAuthn} API.
	%Whenever the webpage registers or authenticates, it calls the attacker's functions instead.
	The attacker can set up a virtual authenticator to register their authenticator instead of the user's~\cite{guan2022formal}.
	%Malicious attackers can also execute this attack by leveraging XSS without a browser extension.
	
	\item \textbf{Modify incoming and outgoing web requests.}
	An attacker can also attempt to compromise the FIDO2 protocol by intercepting HTTP requests and responses.
	Using the data in these requests and responses, the attacker can perform the FIDO2 flow for their authenticator and then send the calculated values directly to the web server from the client's machine.
\end{enumerate}

These attacks map onto the capabilities of the \textit{Malicious DOM Exfiltrator} and \textit{Web Request Exfiltrator} identified in our threat model for password entry (see \S\ref{sec:threat}).

\subsection{Design}

\begin{figure}
	\centering
	\includegraphics[width=\columnwidth]{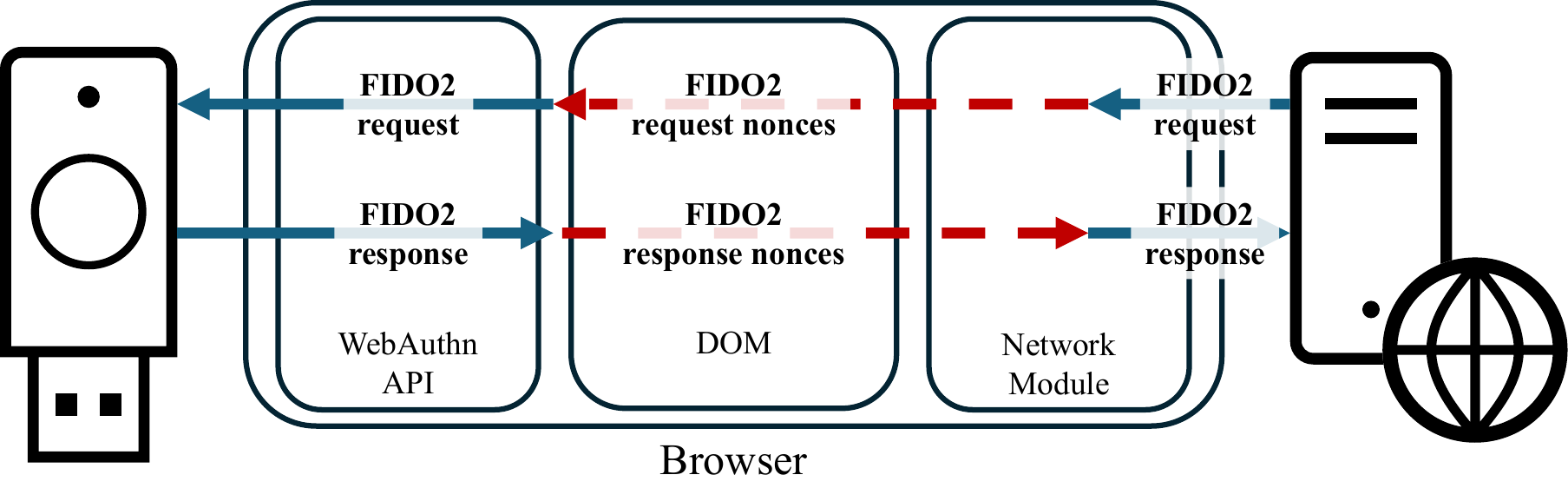}
	\caption{A diagram of our FIDO2 secure browser channel}
	\label{fig:sbc-fido2-browserdesign}
	\Description{Image showing the FIDO2 flow using our prototype. Critical values are never available in the DOM or to malicious extensions, preventing local FIDO2 attacks.}
\end{figure}

To address the above attack vectors and adversaries, we present the design of a secure browser channel for the FIDO2 protocol.
In our design, FIDO2 values shared through this channel are inaccessible to DOM- or extension-based scripts.
At a high level, this channel is created by replacing any sensitive values in an HTTP request with nonces as soon as the browser receives them, long before the DOM or extensions can access them.
The nonces are then only replaced with the actual values before being passed to the browser's internal WebAuthn API, which is neither accessible nor modifiable by the DOM or extensions.
A similar replacement happens for values returned from the the authenticator.
This process is diagrammed by Figure~\ref{fig:sbc-fido2-browserdesign}.
More specifically, this process can be decomposed into the following five steps:

First, when sending a dummy FIDO2 registration or authenticator HTTP response, the webpage (\rp) will replace the FIDO2 payload with dummy values.\footnote{These values can be randomly generated or hard-coded.}
The real FIDO2 payload will be placed inside a \texttt{webauthn\_request} header.
Additionally, \rp sends a \texttt{URL\_resp} header that specifies the URL to which the FIDO2 response should be sent.
Otherwise, the FIDO2 workflow on the server remains unchanged.

Second, when receiving an HTTP response, the browser scans it for the \texttt{webauthn\_request} header.
If detected, the browser will store this value internally and then strip the header from the HTTP request.
Critically, this stripping operation happens before browser extensions can access the web response headers.

Third, the webpage will execute the FIDO2 workflow as normal.
The FIDO2 response contains dummy values precisely so that no change is needed in how the webpage implements FIDO2 using client-side JavaScript.
When this workflow interacts with the FIDO2 \texttt{webAuthn} API, instead of using the values passed, the values originally sent using the \texttt{webauthn\_request} will be used instead.
Importantly, this replacement happens within the browser, not the web page's DOM, preventing access to the real values by scripts running within the webpage (including scripts injected by extensions).

Fourth, when the \texttt{webAuthn} API returns values, these values are intercepted and stored by the browser, then replaced with randomly generated dummy values before being passed off to the webpage script that originally called the \texttt{webAuthn} API.
The browser will also maintain a mapping between the random and real values, allowing them to be replaced later.
Still, as in step 3, scripts on the webpage cannot access the real values (i.e., they are not stored anywhere within the DOM or the webpage's JavaScript execution environment).

Fifth, when an HTTP request is made to the endpoint specified in the original \texttt{URL\_resp} header, the browser will replace the FIDO2 dummy values with the appropriate real FIDO2 response values.
This response will be sent to the website, completing the FIDO2 registration or authenticator process.

To safeguard FIDO2 requests and responses from interference by browser extensions, care must be taken when FIDO2 values are stripped from HTTP response headers or placed in HTTP request payloads.
To this end, response headers must be stripped before \texttt{onHeaderReceived} event, the first chance that browser extensions can access headers in an HTTP response.
Similarly, the real FIDO2 values must be inserted into HTTP requests after the \texttt{onSendHeaders} event, the last chance that browser extensions can read the request body.

\paragraph{Session accountability} 
In addition to the above changes, we modify the FIDO2 popup displayed by the browser to inform the user which account they are authenticating.
This helps provide session accountability.
While not core to the defense described above, it is easy to add with little cost.

\subsection{Implementation}
\label{sec:browser-impl}

To validate our design's feasibility, security, and deployability, we implemented it into Mozilla Firefox 104.0a1.
%To promote open science, this implementation can be found in the Firefox repository (branch names redacted for anonymity).
%
In the modified Firefox, the browser examines every incoming server response in the \texttt{nsHttpChannel}, the earliest point of entry for HTTP packets in Firefox.
Upon detecting a header containing the \texttt{webauthn\_req} field, the browser retrieves the values of \texttt{webauthn\_req} and \texttt{URL\_resp}, passes them to a singleton file and subsequently removes them from the header.
The dummy request in the payload undergoes the standard flow, where browser extensions can interact with it.
The \texttt{WebAuthnTransactionParent} retrieves the original FIDO2 request from the singleton file, creates a \texttt{WebAuthnMakeCredentialInfo} object, and transmits it to the authenticator through the \texttt{WebAuthnManager}'s register function. 
The user interacts with the security token, and the authenticator returns the FIDO2 response.
The original FIDO2 response is stored in the singleton file by the \texttt{WebAuthnTransactionParent}, while a dummy FIDO2 response is provided to the DOM.
When the \texttt{nsHttpChannel} detects an outgoing web request to \texttt{URL\_resp}, it retrieves the original FIDO2 response and adds it to the outgoing request's header.

To accomplish the secure storage, we utilize the singleton design pattern, which facilitates the secure storage and redirection of FIDO2 requests and responses away from browser extensions. The singleton pattern restricts the instantiation of a class to a single instance within a given process. In our design, the singleton class acts as a container for requests and responses, which are stored during the early stage of the flow and later injected back into the flow at the final stage. This design simplifies access to the stored variables, as the singleton file is available anytime during the flow.

Software developers have debated the use of the singleton pattern due to its potential as a global class. However, in our specific implementation, a global class is necessary to establish a direct and efficient path for securely transferring requests and responses from the beginning to the end of a flow. Moreover, the singleton file serves as a global object, eliminating the need to pass object references between different objects and reducing the risk of coding errors.

To ensure our implementation worked as intended, we tested it against a Java-based WebAuthn Demo Server~\cite{java-webauthn}.
Modifying this demo server to support our design only required a small tweak to the code (see Listing~\ref{lst:sbc-webpage}).

To support SBC-FIDO2, we modified this code by changing the response handling as shown in Listing~\ref{lst:sbc-webpage}.

\begin{figure}
\begin{lstlisting}[language=javascript,caption={Web page changes to implement our FIDO2 design},captionpos=b,label=lst:sbc-webpage]
	// Original code
	response.ok(json)
	.build();
	
	// New code
	response.ok(json_dummy)
	.header("webauthn_req", json)
	.header("URL_resp", url)
	.build();
\end{lstlisting}
\Description{Source code demonstrating how to use our FIDO2 defense on websites.}
\end{figure}

\subsection{Security Evaluation}
In our implementation, the original FIDO2 values never reach the DOM, being replaced with nonces long before the DOM could access the values.
Similarly, replacement happens inside browser internals in a location that provides no access to extensions.
As such, our design prevents any unauthorized access or tampering by malicious browser extensions.

To empirically confirm this analysis, we leveraged the work of Yadav and Seamons~\cite{yadav2024security} to craft malicious browser libraries and extensions that seek to break the FIDO2 protocol described in our threat model.
When our defense is not used, these attacks are successful, allowing for one-time malicious logins and for attackers to register their authenticator in place of the users.
After our defense was enabled, both attacks immediately failed.
We also built several variants of these attacks, trying to see if understanding how our implementation worked would allow us to break them, but we never found a way to access the sensitive values from any JavaScript or extension API.

\subsection{Deployability Evaluation}
Regarding deployability, our implementation introduces modifications to the browser architecture to establish a secure channel between websites and the authenticator.
Most specifically, our design necessitates the inclusion of an additional header in the FIDO2 web request and response.
While making this change is a matter of only a couple lines of code (see Listing~\ref{lst:sbc-webpage}), this is still a limitation to the deployability of our approach, especially when compared to our secure browser channel for password entry, which required no changes from the majority of websites.
Still, as the breadth of FIDO2 deployment is somewhat limited, now is the time to make the changes we are proposing.

As a stopgap, future research can explore approaches for automatically identifying FIDO2 values in FIDO2 requests, regardless of how they are included (e.g., in the body, as part of an object).
Doing so could allow our defense to work even without changes to the server.
However, for long-term implementation, we recommend adopting the headers approach due to its minimal overhead and the flexibility it offers servers.
%, granting them the ability to permit extensions to intercept requests if required for specific purposes.
%Additionally, this option to allow interception may prove particularly advantageous for applications beyond the scope of FIDO2, as detailed in Section~\ref{sec:additional-applications}.

\subsection{Implementing in Other Browsers}

To port our FIDO2 defense to other browsers, it is necessary to make the following changes to the codebase:

\begin{enumerate}
	\item Intercept the incoming FIDO2 web response, storing the nonce values and the FIDO2 values.
	\item At the WebAuthn API, the incoming nonce values will be replaced with the FIDO2 values. Also, values returned from the WebAuthn API will be replaced by nonces.
	\item Intercept the outgoing FIDO2 web request, replacing the nonce values with the FIDO2 values from the WebAuthn API.
\end{enumerate}

\paragraph{Chromium}

The incoming FIDO2 web response will be intercepted and modified in the \code{URLLoader::BlockResponseForOrb} method (\code{url\_loader.cc}).
Changes to the WebAuthn API will happen in the \code{AuthenticatorImpl} class (\code{authenticator\_impl.cc}).
The outgoing FIDO2 web request will be intercepted and modified in the \code{URLLoader::OnBeforeSendHeaders} method.
Communication between the network process and the WebAuthn API process will occur using Mojo messaging.

\paragraph{Webkit}

The incoming FIDO2 web response will be intercepted and modified in the \code{NetworkResourceLoader::didReceiveResponse} method (\code{NetworkResourceLoader.cpp}).
Changes to the WebAuthn API will happen in the \code{AuthenticatorGetInfoResponse} class (\code{AuthenticatorGetInfoResponse.cpp}) as part of the CTAP request construction.
The outgoing FIDO2 web request will be intercepted and modified in the \code{NetworkResourceLoader:: continueWillSendRequest} method.
Communication between the two processes will occur using the existing IPC framework.

%% file: discussion.tex
%!TEX root=paper.tex

\section{Discussion}
\label{sec:discussion}

Our design and implementation of secure browser-based authentication channels represent a significant step forward for password managers and 2FA security.
With the goal of merging our APIs into Mozilla Firefox, we frequently consulted with the Mozilla Firefox developers.
They provided feedback on our designs, ensuring they were suitable for use in a modern browser, and reviewed our code on several occasions.
We also followed coding best practices, hoping that this would help speed up the deployment of our APIs in Firefox and other browsers.
%TODO: Add location of both defenses
Our code base is available at \url{https://doi.org/10.5281/zenodo.16739763}, allowing others to replicate and extend our work.

%Lastly, we want to emphasize that implementing our APIs was far from trivial.
%Identifying how to properly design and implement our APIs took hundreds of hours.
%In discussions with the Firefox developers, they often indicated uncertainty about how necessary functionality could be added, particularly regarding functionality that touched multiple parts of the code base and needed to cross trust boundaries.

\subsection{Feedback from Product Teams}
As we developed our defense and prepared this paper, we reached out to product teams to gather their thoughts on our defense.
When reaching out, we provided them with an outline of our defense, a copy of this paper, and offered to answer any questions they had.
As those responding to our query were not making official statements on behalf of their companies, we do not provide direct quotes nor attribute feedback to specific product teams.

Several individuals associated with the browsers respond to our queries.
All were open to the idea of our defense being implemented into the browser.
However, they also all mentioned that resource constraints meant it was unclear when there would be available bandwidth to implement the defense.

In their response, one browser team member mentioned a prior effort to protect credentials from DOM-based adversaries: the Digital Credentials API.\footnote{https://w3c-fedid.github.io/digital-credentials/}
They stated that this effort stumbled due to the requirement for websites to implement the defense and the lack of defense against a reflection attack.
In contrast, our defense requires no changes from the majority of websites and has protections against reflection attacks (\S\ref{sec:reflection}).
Moreover, the digital credentials API does not protect against malicious extensions, which our defense does.
Taken together, the existence of this API demonstrates that browser makers are interested in addressing this problem and that our defense addresses some key challenges that have stymied a prior effort.

We also had a couple of team members from a password manager respond to our queries.
They indicated that one of their customers' key concerns was the ability of malicious extensions to exfiltrate passwords.
The recommendation from this password manager to their customers was to address this by disallowing all extensions other than the password manager.
They were very intrigued by the ability of our defense to resist attack by malicious extensions.
They also recommend that we get this defense standardized through the W3C, as that provides the best pathway for mass adoption.
This is something that we plan to do shortly.

\subsection{Additional Applications}
\label{sec:additional-applications}

In this paper, we demonstrated that our proposed browser-based secure channel could be generalized to protect against local attacks against the FIDO2 protocol.
Similarly, our approach of constructing a browser-based secure channel using nonce replacement can be used to ensure the integrity and confidentiality of data transmitted to and from other browser APIs, protecting against DOM- and extension-based attacks.
We believe the following APIs could also be secured using our high-level design:

\begin{itemize}
	\item The Clipboard API empowers web applications to respond to clipboard commands and seamlessly interact with the system clipboard.
	\item The File System Access API provides read, write, and file management capabilities, while the File API grants access to files and their contents.
	\item The File and Directory Entries API simulates a local file system, enabling web apps to navigate and manipulate files effortlessly.
	\item Geolocation API allows users to share their location with web applications, and the Media Capture and Streams API facilitates seamless audio and video media capture and streaming.
\end{itemize}

%(i) The Clipboard API empowers web applications to respond to clipboard commands and seamlessly interact with the system clipboard. (ii) The File System Access API provides read, write, and file management capabilities, while the File API grants access to files and their contents. (iii) The File and Directory Entries API simulates a local file system, enabling web apps to navigate and manipulate files effortlessly. (iv) Geolocation API allows users to share their location with web applications, and the Media Capture and Streams API facilitates seamless audio and video media capture and streaming.

Furthermore, several experimental browser APIs can be secured from browser extensions without direct JavaScript interaction. These APIs include Web Bluetooth (for Bluetooth Low Energy device connectivity), Barcode Detection (for barcode recognition in images), WebOTP (for phone number verification using one-time passwords), Web NFC (for NFC data exchange), HID (for communication with human input/output devices), WebUSB (for exposing non-standard USB-compatible device services), MediaStream Image Capture (for capturing images and videos), and Contact Picker (for selecting and sharing limited contact details).

Other APIs could benefit from this solution. The key takeaway is that by employing the proposed defense mechanism, web servers can enhance the security of data exchange and protect against unauthorized access or tampering by browser extensions, thereby ensuring the integrity and confidentiality of sensitive information.

\subsection{Future Work}
\label{sec:fw}

Below we describe future work that could further refine the password exfiltration defense described in this paper.

\subsubsection{Improving Compatibility}
\label{sec:fw:compatibility}
We have identified three approaches for improving compatibility.
%Each of these approaches relies on creating a list of non-compliant websites, which as shown by our testing can be done in a fully autonomous manner.
First, password managers could name and shame non-compliant websites.
Second, password managers could implement custom handling for non-compliant websites, replicating the client-side JavaScript that breaks compatibility.
Unfortunately, while both of these approaches are easy to implement, neither scale well to the long-tail of websites.

Instead, we believe that future work should explore automatically lifting relevant client-side JavaScript from login pages, then running this code during \texttt{onRequestCredentials}.
This approach would involve research into (i) how to use program analysis to automatically identify relevant JavaScript (e.g., through taint analysis), (ii) modifying the browser to run this JavaScript in a locked-down environment that cannot exfiltrate the password, but still has sufficient access to a copy of the DOM to process the password, and (iii) updating the \texttt{onRequestCredentials} phase to allow for the outgoing webRequest to be appropriately modified.

\subsubsection{Copy-And-Pasted Passwords}
\label{sec:fw:copy-paste}
The password nonce designs we created can also work with passwords copied from the password manager.
In this case, the password manager would copy a nonce instead of the password and then watch all browser tabs for the nonce to be pasted.
At this point, the defense would work as if that nonce had been autofilled.
If this feature is implemented, users will still need a way to copy raw passwords to enter them outside of the browser or use them as a fallback if nonce replacement fails.
Alternatively, future research could extend nonce replacement directly to the OS, allowing for nonce replacement in apps that rely on the operating system or its libraries to send network traffic---for example, many Microsoft Store apps use the .NET Core.

\subsubsection{Securing Manual Password Entry} 
\label{sec:fw:manual-entry}
Protecting manual password entry (i.e., users typing passwords) is challenging, particularly as an attacker can inject client-side scripts (honest-but-curious entity, DOM attacker) that listen and record keystrokes (i.e., a key logger).
One possible way to secure manual password entry would be to allow the browser to enter a special password entry mode using a conditioned-safe ceremony~\cite{karlof2009conditioned}.
In this mode, the browser would record the user's keystrokes, replacing each stroke with one or more random password nonce characters.
The browser would also prevent any scripts or extensions from recording keystrokes while the password entry mode is active.
After the user enters the password, they would leave the password entry mode.

While this approach could protect manually entered passwords, research would be needed to ensure its effectiveness.
First, care must be taken when selecting the conditioned safe ceremony to ensure that users always enable it when needed.
Second, it will be necessary to design the password entry mode such that it is clear when it is activated~\cite{dhamija2005battle}.
Third, research would be needed to explore how users could be made aware of and encouraged to use this functionality.
Lastly, user studies would be necessary to ensure it is sufficiently usable to encourage users to adopt this functionality.

\subsubsection{Denial of Service for Nonce Injection}
A local attacker could attempt to sidestep our defense by injecting JavaScript that breaks the login page if it suspects a nonce has been autofilled on the page.
In this case, the user may assume that there is something wrong with password autofill.
In response, they may copy-and-paste or manually enter the password, unwittingly offering it up for exfiltration.
To address this, we recommend implementing nonce-based replacement for copy-and-pasted (\S\ref{sec:fw:copy-paste}) and manually entered passwords  \S\ref{sec:fw:manual-entry}).

Even without these defenses, a denial of service attack isn't guaranteed to be effective.
First, it requires identification of password nonces, which for users with generated passwords may be indistinguishable from the actual password.
Second, injecting a denial of service attack for the login page increases the likelihood that the website, browser, or password manager can detect the local attack, allowing for it to be mitigated for all users.
Such detection is even more likely in cases where generated passwords are assumed to be nonces, as this will fully block logging in for the user, triggering a report by that user of unusual behavior to the website.

\subsubsection{Leveraging Nonces for Attack Detection}
In our design, nonces ensure the confidentiality of credentials.
However, it may also be possible to use nonces to detect attacks.
For example, if the browser detects that a nonce is included in a web request body but that no appropriate substitution is made, this indicates a password exfiltration attack.
The browser could then analyze a webpage's DOM for malicious scripts and potentially share this information with websites.
Similarly, it could search for potentially malicious extensions exfiltrating the nonce.
Alternatively, if websites can differentiate between nonces and passwords, this would also allow them to detect that an attack has occurred when they received a nonce.
In either case, the fear of detection may be enough to reduce the likelihood that an attacker will attempt to steal credentials.

In either case, research will be needed to identify how the browser or website can distinguish nonces while also preventing the attacker from doing so (if the attacker realizes they are stealing a nonce, they will not steal it to avoid detection).
Similarly, research will be needed to ensure that any forensic analysis happens in a way that minimizes risks to user privacy.

\subsubsection{User Confusion}
User confusion is a potential issue with nonce injection defenses.
If users view the autofilled password, they may realize it is not their password, especially if they do not use a generated password.
This could lead to confusion.
Although this issue may not be that prevalent because users are less likely to investigate passwords inserted through a password manager, it is still prudent to study users' behavior when they encounter such confusion.

While the shadow DOM is not sufficiently secure to be used to show users their actual password~\cite{ruoti2016messageguard}, the browser could add a similar utility to securely display the real password to the user without allowing that password to be accessed by malicious client-side scripts or extensions.
Future research could also investigate this potential solution.
%Further research is needed to investigate user perception and the potential impact of confusion on the system's overall security before the widespread adoption of nonce-based solutions.

\subsection{Limitations}

Our API does not protect against a compromised OS or browser, which are considered part of the trusted computing base.
Similarly, we did not hide replacement values from \texttt{chrome.devtools.network}.
These tools are meant to aid developers in debugging unexpected behavior, such as misbehaving nonce-replacement.
This is an intentional choice, as having these APIs report the \texttt{requestBody} with a nonce instead of the password would be possible.

Our defense does not protect against attacks that occur after the password leaves the user's computer (e.g., TLS MitM).
We did consider such attacks in our research, but felt that the defense needed to secure against them required too many changes from browsers, websites, and users.
For more details on these threats and defenses, see Appendices~\ref{appx:threat}~and~\ref{appx:design}, respectively.

In these cases, it is important not to let perfection become the enemy of meaningful improvement.
While there are still attacks that can steal passwords, the defense described provides an immediate benefit to the vast majority of password manager users on the vast majority of websites.

%% file: rw.tex
\section{Related Work}

Password managers serve to help users (a) create random, unique passwords, (b) store the user's passwords, and (c) fill in those passwords.
On desktops, password managers are implemented as browser extensions.
The browser does not provide any password management APIs for these extensions to use~\cite{oesch2020that}.
On mobile devices, there is first-party support for password managers, though this support has significant security issues~\cite{oesch2021emperors}.

\subsection{Security}
Password managers have the potential to provide strong security benefits but also have the potential to act as a single point of failure for users' accounts~\cite{stock2014protecting,li2014emperor,silver2014password,oesch2020that,oesch2021emperors}.
In particular, when passwords are autofilled into websites, they are vulnerable to theft by JavaScript and extensions.
In the most alarming case, if the password manager fails to require user interaction before autofilling passwords and allows passwords to be filled into iframes, it opens users to password harvesting attacks that can surreptitiously steal many if not all their passwords~\cite{stock2014protecting,oesch2020that}.
This problem can be even worse when the operating system enforces incorrect behavior, such as in mobile devices~\cite{oesch2021emperors}.
%Similarly, care has to be taken so that other concurrent programs cannot steal sensitive information when it is being processed by the manager, such as when a manager copies information to the system clipboard~\cite{fahl2013hey,mysk2020clipboard}

\subsection{Usability}
Simmons et al.~\cite{simmons2021systematization} systematized password manager use cases.
They found that today's managers poorly support many password manager use cases. Even when supported, the tools are often targeted at experts rather than the lay users they claim to support.

%Huaman et al.~\cite{huaman2021they} investigated integration problems between desktop password managers and websites, finding that managers often struggle to support modern web standards and that websites are highly heterogeneous in their implementations, leading to difficulties.
%Seiler-Hwang et al.~\cite{seilerhwang2019analyzing} conducted a laboratory user study of four smartphone password managers, finding significant usability issues, particularly in the integration of the manager with apps and browsers, with users rating the managers as having barely acceptable usability.

Lyastani et al.~\cite{lyastani2018better} instrumented a password manager to collect telemetry data regarding password manager usage.
Their results show that users underutilized password generation.
This phenomenon was partly explained in research by Oesch et al.~\cite{oesch2022observational} that surveyed users and showed that many users avoid the security-critical functionality of password managers, such as password generation or password audits because they feel these features were too difficult to use.
Instead, they focus on features with the highest usability, such as autofill.

%Work by Karole et al.~\cite{karole2010comparative} and Ciampa et al.~\cite{ciampa2013comparison} have shown that users prefer different password manager implementations, with non-technical users preferring phone and browser-based managers. In contrast, technical users are more likely to prefer a standalone manager.

\subsection{Relation to Our Work}
Research into the security of password managers over the last decade has consistently shown that the autofill process is an all too common element of security vulnerabilities in password managers~\cite{stock2014protecting,li2014emperor,silver2014password,oesch2020that,oesch2021emperors}.
In this paper, we explore how the autofill process can be transformed from a weakness of password managers to one of their strengths, providing benefits not available for manual entry.
Critically, this security benefit is available without changing user behavior, which is not true for other password manager security benefits~\cite{simmons2021systematization,oesch2022observational}.

%% file: conclusion.tex
%!TEX root=paper.tex

\section{Conclusion}
%TODO: Fix up

%There are many avenues for adversaries to exfiltrate passwords: through client-side scripts, the \texttt{webRequest} API, during network transmission, and through phishing.
In this paper, we explore how the password manager can actively protect passwords from theft.
We take a non-implementable protocol described a decade ago by Stock and Johns'~\cite{stock2014protecting} and show how modern browsers can be modified to enable this protocol.
Moreover, we identify a gap in this protocol's threat model and describe how it can be addressed.

To demonstrate the feasibility of these designs, we implement them in Firefox and Bitward.
Using this prototype, we conduct security, functionality, and performance evaluations.
We demonstrate that our tool effectively stops DOM- and extension-based password exfiltration.
We also show that it does so on 97\% of the Alexa top 1000 websites.
This indicates that our approach will provide an immediate and significant security benefit to the vast majority of websites users employ.

Next, we show a secondary advantage of providing a working implementation of nonce-based password replacement working---it enables research on this topic to progress, enabling meaningful efforts towards securing browser-server interactions.
As an example of this possibility, we extend the FIDO2 API to enable it to defend against recently discovered local attacks~\cite{yadav2024security}.
This helps address a security concern in a widely used protocol and demonstrates the utility and flexibility of our high-level design for a secure browser channel using nonce replacement.
We implement our design into Firefox and evaluate it, demonstrating that it empirically blocks local attacks against the FIDO2 protocol.

The code for both of our implementations can be found at \url{https://doi.org/10.5281/zenodo.16739763}.
As such, our work advances our scientific understanding of secure password entry and makes an important practical contribution.
We hope other researchers will build on this artifact to create compelling security features for password managers and password-based authentication. Future research could expand our approach to secure other authentication methods, use nonces to detect and triage vulnerabilities and secure manual password entry.
Similarly, future research can explore securing other browser APIs using our design.
%There is also room for meaningful research into the user-facing aspects related to secure password entry.

%While not perfect, our implementation is working, publicly available code that can already improve the security of autofill on the vast majority of websites.
%For the remaining websites, it is easy to automatically detect compatibility issues and not use our secure autofill API, preventing any functionality regressions.
%As such, our work not only pushes forward our scientific understanding of secure password entry but makes an important practical contribution.

%As such, it can serve as a jumping-off point for additional research into leveraging the password manager to secure password entry.
%In fact, we think there are many interesting potential directions for future research, such as expanding our approach to secure other authentication methods, using nonces to detect and triage vulnerabilities, and securing manual password entry.
%We also think there is room for meaningful research into the user-facing aspects related to secure password entry.

%% file: appx-threats.tex
%!TEX root=paper.tex

\section{Extended Threat Model}
\label{appx:threat}

In this section, we extend the threat model described in \S\ref{sec:threat} to also include remote attacks.

\subsection{Attack Vectors}

We identify an additional location where a remote attacker can exfiltrate a user's password during the password entry process:

\begin{enumerate}[start=3]
	\item \textbf{Exfiltrate it during transmission.}
	If the password is sent to an attacker-controlled server (e.g., phished), it is trivial for the attacker to steal it.
	Alternatively, if the web request containing the password is sent over HTTP (i.e., not encrypted), a network attacker can read it during transmission.
	Similarly, if HTTPS is used, but the TLS connection is compromised by a man-in-the-middle attacker, they can also read that password during transmission.
\end{enumerate}

\subsection{Adversaries}
Based on this additional attack vector, we identify three remote adversaries:

\subsubsection{Network eavesdropper}
This attacker can passively listen to traffic as it passes between the browser and server (attack vector 3).
If the password is sent in the clear (i.e., not using HTTPS), this attacker will be able to steal it.

\subsubsection{TLS Man-in-the-Middle (MitM) Attacker}
This attacker sits between the browser and the server on the network.
Unlike the network eavesdropper, this attacker can modify packets and attempt to man-in-the-middle TLS connections between the browser and server, allowing it to steal credentials transmitted using HTTPS~\cite{oneill2016tls}.

In our threat model, we recognize two variants of this adversary.
The first variant conducts a TLS MitM attack but does not have access to a properly signed and valid leaf certificate, rendering this attack detectable.
The second variant can obtain a properly signed and valid leaf certificate, making this attack more difficult to detect.\footnote{We recognize that this variant can be detected in many cases using defenses such as certificate transparency. However, as these defenses can have false positives, we only consider this variant fully addressed if the defense works, even if the attack passes all implemented checks.}
This attack is partially thwarted if a defense can address only the first variant.

\subsubsection{Phisher}
This attacker has tricked a user into vising a malicious or compromised website and attempts to get the user to submit their password on that website (attack vector 3).
Since this attacker controls the website, they also have full control of the DOM (attacker vector 1). However, as this capability is already covered by the malicious DOM exfiltrator, we focus on just the phishing part of the attack.
While this attacker must convince a user to visit the phishing webpage, we assume this is feasible~\cite{akerlof2015phishing}.

As with the TLS MitM attacker, we recognize two levels of protection.
Partial protection against this attack is achieved if the defense prevents sending passwords to phishers, but users retain the ability to override this protection---for example, by adding the phishing domain to the list of domains associated with the credential.
Full protection is only achieved if the credential is safe from the attacker, regardless of user action.

%% file: appx-design.tex
%!TEX root=paper.tex

\section{Design Space Exploration}
\label{appx:design}

\input{design-table}

Using our threat model, we explored the design space for password entry defenses.
This exploration was conducted using a mixture of literature review and in-group discussion.
Ultimately, we identified two paradigms for securing passwords from exfiltration: (i) moving password-based authentication into the browser's non-webpage UI and (ii) leveraging a nonce-based replacement scheme for autofilled passwords.
We then describe five designs of our creation that can be used to realize these paradigms, with each design having different trade-offs in terms of security and deployability.
Table~\ref{tab:properties} gives a summary of these designs.

When evaluating the security of designs, we consider which of the six adversaries from our threat model the design would thwart.
When evaluating the deployability of designs, we consider the burden that would be placed on three entities:

% TODO: This could use some good citations.
\begin{enumerate}
	\item \textbf{Users.}
	Changing user workflows could cause confusion and lead users to turn off the defense~\cite{herley2009so}.
	Even small changes can challenge users, particularly the elderly~\cite{das2020non}.
	
	\item \textbf{Websites.}
	Solutions requiring millions of websites to update their authentication flows will likely see slow adoption~\cite{bonneau2012quest}.
	This can be seen with the adoption of 2FA and passkeys, which while having been adopted by larger websites and SSO providers are still not supported by the vast majority of websites.
	
	\item \textbf{Browser vendors.}
	The number of browsers is far smaller than the number of users and websites on the Internet.
	As such, solutions that only require changes to the browser are preferable, as they will require the fewest entities to adopt them.
	Still, modern browsers are extremely complex, and as such, changes to browsers are non-trivial.
	
\end{enumerate}

We do not consider the burden on password manager vendors, as they are responsible for updating their tools to provide as much password security as possible.

\subsection{Baseline}

To provide a baseline, we first discuss the security and usability of the password autofill workflow (see \S\ref{sec:workflow}).

\paragraph{Security}
After autofill, the password is stored in the page's DOM, leaving it vulnerable to all three local adversaries.
Additionally, while password managers may check security indicators related to the webpage where autofill occurs~\cite{oesch2020that}, they do not generally check the security of the outgoing connection (i.e., they do not have \texttt{webRequest} API access) and are vulnerable to passwords being sent over unsecured or improperly secured connections.
Finally, since password managers only list passwords associated with the current domain, they provide some protection against phishing attacks~\cite{oesch2021emperors}, as no passwords will be available for autofill on a phishing domain.
Still, as users can override this behavior by associating the phishing domain with a legitimate password, we rate this as only partially stopping a phishing attack.

\paragraph{Deployability}
As this is the baseline, by definition, we consider it to require no changes to user workflows, websites, or browsers.
For users, we recognize that adopting a password manager requires a change in workflow, but we consider this cost to have already been paid.
For websites, we note that there can be interaction issues between websites and password managers~\cite{huaman2021case}. However, as these can be addressed through additional (potentially targeted) engineering of the password manager, we still count this as requiring no website changes.
For browsers, password managers can work as an extension, requiring no further changes from the browser (though some browsers do provide an integrated password manager).

\subsection{Paradigm: In-Chrome Password Entry}

The first paradigm we identify involves moving the authentication workflow from the webpage into the browser's non-webpage UI~\cite{ruoti2017end}.
In this paradigm, the webpage would indicate to the browser that the user needs to enter their password.
The browser would then handle password entry and transmission within its non-webpage UI (i.e., outside of the DOM), effectively preventing access to the password for local attackers.

We identify two possible designs for realizing this paradigm.

\subsubsection{Design \#1: Browser-Managed Authentication}

To implement this paradigm, our first design leverages the \texttt{WWW-Authenticate} and \texttt{Authorization} HTTP authentication headers.
These headers are used to trigger an authentication workflow in the browser.
Most commonly, this is the \texttt{basic} authentication workflow, which shows the users a popup where they can enter their password.

Still, several changes to the design of a browser's HTTP authentication handling would be necessary before it could suitably replace in-page authentication.
First, a new interface is needed that is more phishing resilient.
The current interfaces are displayed on top of the webpage, leaving them vulnerable to spoofing attacks~\cite{bravo2012operating}.
To solve this issue, new UI elements would be introduced away from the webpage where password entry would take place~\cite{ruoti2017end}.
Second, integration with password managers would need to be supported, as currently, password managers cannot recommend or autofill credentials in the HTTP authentication dialogs.

Finally, we advocate for adding the ability to respond to a \texttt{WWW-Authenticate} header not with an \texttt{Authorization} header but rather with a standard HTTP request containing the password sent to an endpoint specified in the \texttt{WWW-Authenticate} header.
This would obviate the need for websites to add code to process a new header, instead allowing them to use their existing code with little to no modification other than adding a \texttt{WWW-Authenticate} header to the login page.

\paragraph{Security}
As password entry and processing happens outside of the DOM, the password is unavailable to any local adversaries.
Additionally, as the HTTP request with the password is managed by this design, it can entirely refuse to send it over an HTTP or bad HTTPS connection.
Still, a TLS MitM attacker with appropriate certificates will appear legitimate, so this design does not protect against this threat.
Lastly, it provides no additional phishing protection over the baseline, as users can still associate phishing domains with credentials.

\paragraph{Deployability}
The new in-chrome password entry UI will require users to learn a new workflow, significantly limiting its deployability.
This is even more true when it is recognized that users need to learn the workflow well enough to avoid spoofing attacks~\cite{bravo2012operating,ruoti2017end}.
This design will require website changes, though since it allows submission of credentials to existing endpoints, the change is as simple as adding a header to a login page.
Since this is a minimal change, we rate it as having partial support for not requiring significant website changes.
Lastly, this will require significant changes to the browser, both in terms of new UI elements and backend processing.

\subsubsection{Design \#2: Strong Password Protocols}
Design \#1 can be further enhanced by incorporating a strong password protocol~\cite{wu1998secure,jarecki2018opaque}.
These protocols have the benefit that passwords are never sent over the wire, with knowledge of the password demonstrated using a zero-knowledge proof, proving protection against network attacks and phishing.
One of the hurdles to implementing strong password protocols is requiring trusted password entry interfaces to be secure~\cite{ruoti2017end}, but Design \#1 already provides this.

\paragraph{Security}
As with Design \#1, the password is never present in the DOM and is thus unable to be exfiltrated by local attackers.
Additionally, strong password protocols do not require a secure connection, meaning they are secure against all attackers, including attackers who can transparently MitM TLS connections.
Lastly, even if a strong password protocol is conducted with a phisher, that phisher will not learn anything about the password.

\paragraph{Deployability}
This design has all the limitations of Design \#1.
Moreover, it requires a complete reworking of the website authentication flow, significantly limiting its chances of adoption.\footnote{Strong password protocols have existed for decades but have never seen widespread adoption~\cite{hao2022sok}.}
Browsers would also need to support the selected zero-knowledge protocols.

\subsection{Paradigm: Password Nonce}

Nearly a decade ago, Stock and Johns~\cite{stock2014protecting} evaluated the security of password manager autofill, finding that managers autofilled passwords into malicious websites under many conditions.
At the end of their paper, they proposed a possible solution to this problem:

\begin{enumerate}
	\item When an autofill operation is triggered, a random placeholder for the password (\emph{a password nonce}) is generated.
	\item The password nonce is autofilled into the webpage.
	\item The password manager will scan outgoing \texttt{POST} requests looking for the password nonce.
	\item If the password nonce is detected while being transmitted to an origin that matches the origin for the real password, the manager will replace the nonce with the password.
\end{enumerate}

While the design proposed by Stock and Johns is incompatible with modern browsers, the core paradigm remains sound.
As such, we create three designs for implementing this paradigm that are compatible with modern browsers.
We also evaluate these designs against a stronger threat model than used by Stock and Johns, which didn't consider or protect against malicious browser extensions with web request access.

In comparison to the first paradigm, password nonces benefit from requiring no changes to user workflow---i.e., it is identical to using a password manager normally.
There is also no need to change websites.

\subsubsection{Design \#3: JavaScript-Based Nonce Injection}
For this design, we craft a solution that works entirely in the DOM, requiring no changes to the browser.
To accomplish this, the password manager (which has full access to the DOM) will take the following steps:

\begin{enumerate}
	\item If the password form already has an \texttt{onsubmit} method, the manager will store this method. It will also store any \texttt{submit} event listeners.
	\item The password manager will remove the \texttt{onsubmit} method or any \texttt{submit} event listeners from the form. If, in the future, a new \texttt{onsubmit} method or \texttt{submit} listener were to be added, this operation will be prevented, with the method or listener replacing or being added to the stored items, respectively.
	\item The password manager will set the \texttt{onsubmit} method for the form. The new \texttt{onsubmit} method will perform the following operations:
	\begin{enumerate}
		\item Call the stored \texttt{onsubmit} method, if any.
		\item Call any stored submit event listeners.
		\item Replace the password nonce with the actual password as long as the form data will be submitted to an appropriate origin.
		\item Submit the form.
		\item Replace the password with the nonce (in case the submit operation does not cause a whole-page refresh).
	\end{enumerate}
\end{enumerate}

As the password is replaced immediately preceding the submit operation, there should be no chance for in-page JavaScript to read this value.
Items 1--2 are necessary to ensure that no other scripts run after the password replaces the nonce.

\paragraph{Security}
\begin{figure}
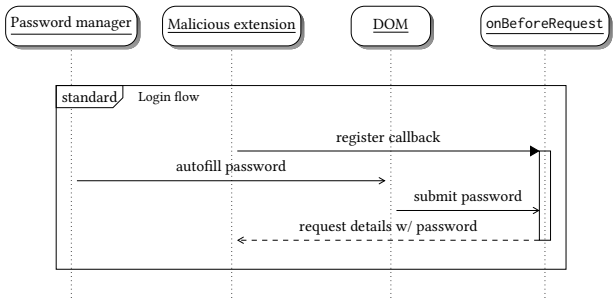

	\centering
	\adjustbox{max width=\columnwidth}{
		\begin{sequencediagram}
			\tikzset{inststyle/.append style={
					drop shadow={top color=gray, bottom color=white}, 
					rounded corners=2.0ex
				}
			}   
			
			\newinst{p}{Password manager}
			\newinst[0.5]{m}{Malicious extension}
			\newinst[1]{d}{DOM}
			
			\newinst[1]{b}{\texttt{onBeforeRequest}}
			
			\begin {sdblock}{ standard }{Login flow}
			\begin{call}{m}{register callback}{b}{request details w/ password}
				\mess{p}{autofill password}{d}
				\mess{d}{submit password}{b}
			\end{call}
		\end{sdblock}
	\end{sequencediagram}
}

\caption{Diagram illustrating how an attacker can use an \texttt{onBeforeRequest} callback to exfiltrate passwords.}
\label{fig:exfiltrate}
\end{figure}

In theory, this design prevents against DOM-based exfiltration.
However, a malicious adversary (i.e., not the honest-but-curious DOM observer) could attempt to circumvent this protection in various ways.
First, they could attempt to replace the original form with a look-a-like form that does not trigger this design, allowing for direct exfiltration of this password.
Second, they could look for other events that get called between replacing the nonce and the submit event.
While we are unaware of any such events, that does not mean new ones will not be added in the future, an inherent problem in relying on non-security JavaScript functionality for security~\cite{ruoti2016messageguard}.
At best, this would lead to a cat-and-mouse game between attackers and defenders, which is not desirable.
Moreover, exfiltration is trivial for a web request exfiltrator, as the nonce is replaced by the password before the submit event, leaving the password within the web request (see Figure~\ref{fig:exfiltrate}).

The password manager can attempt to prevent submission to non-HTTPS sites by examining the \texttt{action} property prior to nonce replacements and detecting and reversing any changes to this property after submission.
However, this is once again a case of trusting non-security JavaScript functionality for security, which is dangerous at best.
Also, as this defense exists solely within the DOM, it cannot check the security of the outgoing connection to prevent bad TLS connections.
Lastly, it provides no more protection against phishers than the base password autofill workflow.

\paragraph{Deployability}
The password manager implements this approach fully within the DOM, so it requires no changes to user workflows, websites, or browsers.
%Still, like Design \#2, this approach could break websites that rely on scripts to submit the password instead of the form.

\subsubsection{Design \#4: API-Based Nonce Injection}

The security limitations of Design \#3 arise from operating within the DOM and relying on non-security JavaScript functionality to implement secure code.
In Design \#4, we address those issues by modifying the browser extension \texttt{webRequest} API, moving the nonce replacement into this API so that it occurs after the form is submitted and in a location where no JavaScript will have access to the replaced password.

In this design, when the password manager autofills the password nonce, it will also register a nonce replacement callback with \texttt{webRequest} API.
After the form is submitted and immediately before the \texttt{onBeforeRequest} event in the \texttt{webRequest} API (see Figure~\ref{fig:webrequest}), the password manager will call the registered replacement callback, with the callback returning one or more objects contains the following items:
\begin{enumerate*}[label=\emph{(\roman*)}, itemjoin={{, }}, ]
\item the name of the field that was autofilled with the password nonce
\item the password nonce
\item the replacement password
\item the origin (scheme, host, port) associated with the password
\end{enumerate*}.
When deciding what to return, the password manager has full access to the current request (headers and body), ensuring that the request passes all relevant security checks (e.g., checking for proper destination) before the replacement password is provided.

The browser will then search through the key-value pairs in the \texttt{FormSubmission} object (see \S\ref{sec:formsubmission}), making the requested substitution if and only if
\begin{enumerate*}[label=\emph{(\alph*)}, itemjoin={{, }}, ]
\item the indicated field's value matches the password nonce exactly
\item the web request will be sent to the indicated origin
\end{enumerate*}.
%Requiring (a) may lead to some compatibility issues if a script creates a web request using a different field name, though we did not encounter this in our testing (see \S\ref{sec:evaluation}).
%If necessary, this requirement could be dropped, though it might have unexpected consequences.\footnote{We could not identify any security issues that couldn't already be caused by malicious client-side scripts.}
Because replacement happens within the \texttt{FormSubmission} object, proper sanitization and encoding of the key-value pairs is handled by the browser.
From this point on, the browser processes the web request as normal.

A major benefit of doing the replacement immediately before \texttt{onBeforeRequest} is that the web request has not already been serialized, making it relatively easy to update and simplifying the implementation of this design in browsers.
Additionally, the browser code that needs to be modified is all in a single location.

%Finally, we note that unlike Design \#3, this design allows copying and pasting of password nonces from the password manager.
%As the web request API examines all outgoing web requests, replacement of nonces is still possible.
%Still, there will need to be functionality allowing users to copy real passwords if they are to be entered outside of the browser (though this can be made harder to activate, incentivizing the use of nonces where possible).

\paragraph{Security}
The password is not vulnerable to DOM-based attackers because the nonce replacement occurs outside of the DOM.
However, as the replacement happens early in the \texttt{webRequest} workflow, the password will be visible in later stages of the workflow, allowing it to be exfiltrated by malicious or compromised browser extensions with \texttt{webRequest} permissions (see Figure~\ref{fig:exfiltrate}).

In contrast to Design \#3, the browser can examine the outgoing connection and ensure that it properly matches, refusing to autofill for HTTP and bad HTTPS connections.
Still, a TLS MitM attacker with access to a valid leaf certificate can exfiltrate the password.
Phishing will also be stopped (as the origin is different) unless the user associates the phishing domain with the password.

\paragraph{Deployability}
As with Design \#3, no changes to the user workflow or websites are required.
However, changes to the browser are required.
Still, as mentioned above, the benefit of doing the replacement immediately before the \texttt{onBeforeRequest} step is that the web request is still mutable at this time, limiting the size of the changes needed in the browser.
For this reason, we say this property is still partially achieved.

\subsubsection{Design \#5: Browser-Based Nonce Injection}\label{sec:design5}

To build on the security of Design \#4, this design moves the replacement step even later in the \texttt{webRequest} workflow, adding an \texttt{onReplaceCredential} stage that occurs immediately after \texttt{onSendHeaders}.
This is the last stage where the \texttt{webRequest} can be modified before sending it.
Additionally, at this point, extensions can no longer see the body of the current web request, preventing them from seeing the replaced passwords.
Otherwise, this solution works the same as Design \#4, with the callback providing details about the web request as it was prior to any nonce replacements.

The major drawback with Design \#5 is that the \texttt{webRequest} API was not designed to support modifications to the web request body or communication with extensions at this point.
This has two major implications.
First, the web request body must be deserialized, modified, and reserailzed (with appropriate sanitization and encoding).
This has a non-negligible performance cost (see \S\ref{sec:overhead-evaluation}), though it only needs to be paid when replacements are made.
Second, this approach requires changes to the browser in multiple locations in the codebase.
While we were able to overcome this issue in our implementation (see \S\ref{sec:implementation}), it still does mean that implementing this change across all browsers will take more effort than implementing Design \#4.

\paragraph{Security}
As with Design \#4, Design \#5 is impervious from DOM-based password exfiltration.
Additionally, since replacement happens after extensions can view the request body, it also resists exfiltration by browser extensions.
Regarding remote attackers, it has the same strengths and weakness as Design \#4.

\paragraph{Deployability}
This design has largely the same deployability characteristics as Design \#4, except that it requires changes to the browser in multiple locations.

\subsection{Selecting a Design}

Examining the five designs, we see that all five designs improve upon the security of the current password entry process used by password managers.
Design \#2 (strong password protocols) has the best security.
However, as it requires a complete redesign of the authentication workflow for users, websites, and browsers, it is unlikely that this design will ever see widespread adoption.

Looking at the remaining designs, we see that Designs \#1 and \#5 offer the same level of security, with Design \#5 having a much higher deployability score.
As such, we find Design \#5 to be the most compelling design from our design space exploration.

%% file: design-table.tex
\begin{table}
	\setupratingstable
	\adjustbox{max width=\columnwidth}{
		\begin{tabular}{ll|lll|lll|lll|}
			\cline{3-11}
			& & \multicolumn{6}{c|}{Security} & \multicolumn{3}{c|}{Deploy}
			\\ \cline{3-11}
			& Design	 					
			& \rowheader{Protection from DOM observer}
			& \rowheader{Protection from DOM exfiltrator}
			& \rowheader{Protection from web request exfiltrator~}
			
			& \rowheader{Protection from network observer}
			& \rowheader{Protection from TLS MitM attacker}
			& \rowheader{Protection from phisher}
			
			& \rowheader{No changes to user workflow}
			& \rowheader{No changes to websites}
			& \rowheader{No changes to browsers}
			\\ \hline
			
			& Current password manager autofill
			& \snone	& \snone	& \snone
			& \snone	& \snone	& \spart
			& \sfull	& \sfull	& \sfull
			\\ \hline
			
			1. & Browser-managed authentication
			& \sfull	& \sfull	& \sfull
			& \sfull	& \spart 	& \spart
			& \snone	& \spart	& \snone
			\\
			
			2. & Strong password protocols
			& \sfull	& \sfull	& \sfull
			& \sfull	& \sfull	& \sfull	
			& \snone	& \snone	& \snone
			\\ \hline
			
			3. & DOM-based nonce injection		
			& \sfull	& \snone	& \snone
			& \snone	& \snone	& \spart
			& \sfull	& \sfull	& \sfull
			\\	
			
			4. & Extension-based nonce injection		
			& \sfull	& \sfull	& \snone
			& \sfull	& \spart	& \spart	
			& \sfull	& \sfull	& \spart
			\\		
			
			5. & Browser-based nonce injection	
			& \sfull	& \sfull	& \sfull
			& \sfull	& \spart	& \spart
			& \sfull	& \sfull	& \snone
			\\ \hline
			
%			\hline
%			% 2FA for comparison								
%			& Code-based 2FA	
%			& \snone	& \snone	& \snone	& \snone	& \snone
%			& \snone	& \snone	& \sfull
%			\\
%			
%			& FIDO2	
%			& \sfull	& \sfull	& \sfull	& \sfull	& \sfull
%			& \snone	& \snone	& \snone
%			\\ \hline
			
		\end{tabular}	
	}
	
	\vspace{.5\baselineskip}
	\begin{tabular}{ll}
		\sfull & Fully achieves the property \\
		\spart & Achieves the property with some limitations \\
		\snone & Fails to achieve the property \\
	\end{tabular}
	
	\caption{An evaluation of the five approaches to securing password entry.}
	\label{tab:properties}
\end{table}